\documentclass[10pt]{article}
\usepackage{amsmath}
\usepackage{cite} 
\usepackage{indentfirst}
\usepackage{latexsym}
\usepackage{pstricks}

\long\def\ca#1\cb{} 

\newcommand{\ad}{^\dagger }

\newcommand{\AND}{{\small AND}}

\newcommand{\becs}{\begin{cases}}
\newcommand{\bem}{\begin{matrix}}

\newcommand{\colo}{\,\hbox{:}\,}

\newcommand{\dya}[1]{|#1\rangle\langle#1|}

\newcommand{\encs}{\end{cases}}
\newcommand{\enm}{\end{matrix}}

\newcommand{\inpd}[2]{\langle#1|#2\rangle }
\newcommand{\ket}[1]{|#1\rangle }

\newcommand{\lra}{\leftrightarrow }

\newcommand{\msk}{\medskip }

\newcommand{\mted}[3]{\langle#1|#2|#3\rangle }

\newcommand{\NOT}{{\small NOT}}
\newcommand{\od}{\odot }
\newcommand{\OR}{{\small OR}}
\newcommand{\ot}{\otimes }

\newcommand{\ra}{\rightarrow }

\newcommand{\st}{\sqrt{2}}

\newcommand{\Tr}{{\rm Tr}}

\newcommand{\vb}{\,|\,}

\newcommand{\DM}{{\mathcal D}}

\newcommand{\FM}{{\mathcal F}}

\newcommand{\HM}{{\mathcal H}}

\newcommand{\PM}{{\mathcal P}}
\newcommand{\QM}{{\mathcal Q}}

\newcommand{\al}{\alpha }
\newcommand{\bt}{\beta }
\newcommand{\gm}{\gamma }

\newcommand{\dl}{\delta }

\ca
\def\outl#1{\par{\medskip\noindent\hspace*{.5cm}\bf
      \mathversion{bold}#1\mathversion{normal}\smallskip} }
 \def\xa{} \def\xb{}  
\cb

 \def\outl#1{}  \def\xa{} \def\xb{}  

\ca
 \def\outl#1{\par{\medskip\noindent\hspace*{.5cm}\bf
      \mathversion{bold}#1\mathversion{normal}\smallskip} }
 \long\def\xa#1\xb{}
 
\cb

\begin{document}

\centerline{\Large A Consistent Quantum Ontology}
\xa
\msk

\centerline{Robert B. Griffiths}
\centerline{Physics Department}
\centerline{Carnegie-Mellon University}
\vspace{.2cm}

\centerline{Version of 11 February 2013}
\vspace{0.5 cm}

\xb

\xa
\begin{abstract}

  The (consistent or decoherent) histories interpretation provides a
  consistent realistic ontology for quantum mechanics, based on two main
  ideas.  First, a logic (system of reasoning) is employed which is compatible
  with the Hilbert-space structure of quantum mechanics as understood by von
  Neumann: quantum properties and their negations correspond to subspaces and
  their orthogonal complements.  It employs a special (single framework)
  syntactical rule to construct meaningful quantum expressions, quite
  different from the quantum logic of Birkhoff and von Neumann.  Second,
  quantum time development is treated as an inherently stochastic process
  under all circumstances, not just when measurements take place.  The
  time-dependent Schr\"odinger equation provides probabilities, not a
  deterministic time development of the world.

  The resulting interpretive framework has no measurement problem and can be
  used to analyze in quantum terms what is going on before, after, and during
  physical preparation and measurement processes. In particular, appropriate
  measurements can reveal quantum properties possessed by the measured system
  before the measurement took place.  There are no mysterious
  superluminal influences: quantum systems satisfy an appropriate form of
  Einstein locality.

  This ontology provides a satisfactory foundation for quantum information
  theory, since it supplies definite answers as to what the information is
  about.  The formalism of classical (Shannon) information theory applies
  without change in suitable quantum contexts, and this suggests the way in
  which quantum information theory extends beyond its classical counterpart.
 
\end{abstract}
\xb




	\section{Introduction}
\label{sct1}
\xa

\xb\outl{Scientific advances change our view of the world
}\xa

\xb\outl{Philosophy of science should work out implications
}\xa

Scientific advances can significantly change our view of what the world is
like, and one of the tasks of the philosophy of science is to take successful
theories and tease out of them their broader implications for the nature of
reality.  Quantum mechanics, one of the most significant advances of twentieth
century physics, is an obvious candidate for this task, but up till now efforts
to understand its broader implications have been less successful than might
have been hoped.  The interpretation of quantum theory found in textbooks,
which comes as close as anything to defining ``standard'' quantum mechanics, is
widely regarded as quite unsatisfactory.  Among philosophers of science this
opinion is almost universal, and among practicing physicists it is
widespread. It is but a slight exaggeration to say that the only physicists who
are content with quantum theory as found in current textbooks are those who
have never given the matter much thought, or at least have never had to teach
the introductory course to questioning students who have not yet learned to
``shut up and calculate!''

\xb\outl{In QM task hindered by measurement problem
}\xa

On all sides it is acknowledged that the major difficulty is the \emph{quantum
  measurement problem}.  Significantly, it occupies the very last chapter of
von Neumann's 1932 \textit{Mathematical Foundations of Quantum Mechanics}
\cite{vNmn32b}, and forms what many regard as the least satisfactory feature
of this monumental work, the great-grandfather of current textbooks. The
difficulties in the way of using measurement as a fundamental component of
quantum theory were summed up by Wigner in 1963 \cite{Wgnr63}, and confirmed
by much later work; see, e.g., the careful analysis by Mittelstaedt
\cite{Mttl98}.  A more recent review by Wallace \cite{Wllc08} testifies both
to the continuing centrality of the measurement problem for the philosophy of
quantum mechanics, and to the continued lack of progress in resolving it; all
that has changed is the number and variety of unsatisfactory solutions.

\xb\outl{Two measurement problems.  First = macro Qm pointer superposition
}\xa

Actually there are two distinct measurement problems.  The \emph{first
  measurement problem}, widely studied in quantum foundations, comes about
because if the time development of the measurement apparatus (and its
environment, etc.)  is treated quantum mechanically by integrating
Schr\"odinger's equation, the result will typically be a \emph{macroscopic
  quantum superposition} or ``Schr\"odinger cat'' in which the apparatus
pointer---we shall continue to use this outdated but picturesque
language---does not have a definite position, so the experiment has no
definite outcome.  Contrary to the belief of experimental physicists.

\xb\outl{Second problem = how pointer related to earlier microscopic situation
}\xa

If this first problem can be solved by getting the wiggling pointer to
collapse down into some particular direction, one arrives at the \emph{second
  measurement problem}: how is the pointer position related to the
\emph{earlier} microscopic situation which the apparatus was designed to
measure, and which the experimental physicist believes actually caused the
pointer to point this way and not that way?  When one hears experimental
particle physicists give talks, it sounds as if they believe their detectors
are triggered by the passage of real particles zipping along at enormous speed
and producing electrical pulses by ionizing the matter through which they
pass.  No mention of the sudden collapse of (the modern electronic counterpart
of) the apparatus pointer at the end of the measurement process.  Instead,
these physicists believe not only that the apparatus is in a well-defined
macroscopic state---each bit in the memory is either 0 or 1---after the
measurement, but in addition that this outcome is well correlated with a prior
state of affairs: one can be quite confident that a negative muon was moving
along some specified path at a particular moment in time.  Have they forgotten
what they learned in their first course in quantum theory?

\xb\outl{Experimenter wants to know prior state of system.  Not vN's model
}\xa

It this connection note that laboratory measurements in particle physics are,
in the vast majority of cases, not appropriately modeled by the scheme
proposed by von Neumann in which a seemingly arbitrary wave function collapse
leads to a correlation between the measurement outcome and a property of the
measured system at a time \emph{after} the measurement has taken place.
In a typical particle physics experiment the particle will either have
disintegrated, escaped from the apparatus, or been absorbed long before the
measuring apparatus has registered its behavior, and in any case the
experimentalist is interested in what was going on \emph{before} detection
rather than afterwards.

\xb\outl{Preparation, measurement in Qm orthodoxy
}\xa

The absence of a satisfactory solution to this second measurement problem has
led to the development of a certain quantum orthodoxy which affirms
that the only task of quantum theory is to relate the \emph{outcome} of a
macroscopic \emph{measurement} to an earlier and equally macroscopic (i.e.,
describable in ordinary language) \emph{preparation}.%
\footnote{See, for example, \cite{dMyn02}.} %
The quantum wave function $\ket{\psi(t)}$ associated with the prepared system,
regarded as isolated from its environment until it interacts with the
measuring apparatus (the uniwave in the terminology of Sec.~\ref{sbct3.2}
below) does not have any ontological reference; it is simply a mathematical
device for calculating the probabilities of measurement outcomes in terms of
the earlier preparation procedure. Stated slightly differently, it provides an
abstract representation of the preparation process, or at least as much of, or
that part of, the preparation process which is needed to calculate
probabilities of outcomes of future measurements.

\xb\outl{Black box separates preparation, measurement
}\xa

In this approach ``preparation'' and ``measurement'' can be thought of as part
of the real world, but what happens in between occurs inside a ``black box''
which cannot be opened for further analysis and is completely outside the
purview of quantum theory---although some future theory, as yet unknown, might
allow a more precise description.  One can be sympathetic with strict
orthodoxy in that it is intended to keep the unwary out of trouble.  Careless
thinkers who dare open the black box will fall into the quantum foundations
Swamp, where they risk being consumed by the Great Smoky Dragon, driven insane
by the Paradoxes, or allured by the siren call of Passion at a Distance into
subservience to Nonlocal Influences.  Young scientists and philosophers who do
not heed the admonitions of their elders will, like the children in one of
Grimms' fairy tales, have to learn the truth by bitter experience.

\xb\outl{Measurement problems have also hindered QM research.  Qm information
}\xa

The measurement problems and the associated lack of a clear conceptual
foundation for quantum theory have not only been a stumbling block in quantum
foundations work. They have also slowed down, though fortunately not stopped,
mainstream physics research.  In older fields such as scattering theory, the
pioneers spent a significant amount of time working through conceptual
issues. But once the accepted formulas are in place and yield results
consistent with experiment, their intellectual descendents have the luxury of
calculating without having to rethink the issues which confused their
predecessors.  In fields in which quantum techniques are applied in fresh ways
to new problems, such as quantum information (the technical specialty of the
author of this paper), conceptual issues that have not been resolved give rise
to confusion and wasted time.  Both students and researchers would benefit
from having the rather formal approach to measurements found in, e.g., Nielsen
and Chuang \cite{NlCh00}, to mention one of the best known books on the
subject, replaced by something which is clearer and more closely tied to the
physical intuition needed to guide good research, even in a field heavily
larded with mathematical formalism.  Black boxes can be a useful approach to a
problem, but can also stand in the way of a good physical understanding.

\xb\outl{Qm info as key to Qm foundations has failed
}\xa

At one time it was optimistically supposed that quantum information would
provide a new key to resolving the problems of quantum foundations
\cite{BbFc03,Fchs03}.  However, later developments have not confirmed this
earlier optimism, and Timpson's 2008 review \cite{Tmps08} and his more recent
\cite{Tmps10} provide a clear indication of where the trouble lies. Bell's
question, ``Information about what?'' \cite{Bll90}, has not been answered.
And why not?  Timpson realizes that quantum information cannot simply be about
outcomes of measurements (assuming the first measurement problem has been
solved), for this fails to connect these outcomes with properties of the
system being measured.  And he rejects the idea that measurements can reveal
microscopic quantum properties, for this leads, in his opinion, to hidden
variables and all the insuperable difficulties associated therewith.  Clearly
the problem is a lack of a suitable quantum ontology, something which quantum
information could be about.  (The possibility that quantum information could
be about nothing at all is also discussed by Timpson \cite{Tmps10} under the
heading of ``immaterialism,'' which he does not find satisfactory. For
comments on his own proposal in \cite{Tmps08} for defining quantum information
see Sec.~\ref{sbct7.2}.)

\xb\outl{Thesis of paper: consistent Qm ontology builds on logic, stochastic dynamics
}\xa

\xb\outl{Histories approach from 1984 to CQT
}\xa

The thesis of this paper is that both measurement problems can be, and in fact
have been, resolved: the motion of the pointer stilled and the black box
opened, by a consistent quantum ontology that builds upon two central ideas.
The first is a system of logic that addresses the question of how to reason
about a quantum system described mathematically by a Hilbert space.  The
second is a system of stochastic or random dynamics that applies to all
quantum dynamical processes, not just measurements.  These ideas were brought
together for the first time in the author's ``consistent histories''
interpretation \cite{Grff84}.  Subsequently they were developed by Omn\`es,
whose work has appeared in numerous papers and two books \cite{Omns94,Omns99},
and further developed, to some extent independently, by Gell-Mann and Hartle
\cite{GMHr90} using the name ``decoherent histories.''  The differences
between decoherent and consistent histories are not sufficient (in the
author's opinion) to merit separate discussions, so the single term
``histories'' will be employed below; anyone who disagrees is welcome to
prepend ``consistent'' wherever desired.  The most complete discussion of
histories ideas currently available is the author's \cite{Grff02d}, hereafter
referred to as CQT; for more compact treatments see \cite{Grff09b, Hhnb10} and
the first part of \cite{Grff11b}.%
\footnote{It is somewhat unfortunate that the discussion of consistent
  histories presented in \cite{Wllc08} bears little resemblance to what is
  found in \cite{Grff02d}, despite the latter being listed in the
  bibliography.} %

\xb\outl{Criticisms of histories by Dowker, Kent, et al. Responses ignored by Wallace
}\xa

As is often the case with new ideas, the histories approach was subject to
serious criticisms by (among others) d'Espagnat \cite{dEsp87,dEsp90}, Dowker
and Kent \cite{DwKn96}, Kent \cite{Knt96,Knt97,Knt98}, and Bassi and Ghirardi
\cite{BsGh99,BsGh00}, during the decade and a half that followed the original
publications.  Responses were published in
\cite{Grff98,GrHr98,Grff00,Grff00b}; in some cases a further reply to the
response will be found immediately after the response.  While these criticisms
were (in the author's opinion) largely based upon misunderstandings of the
histories program, they had the good effect of leading to a better and clearer
formulation of its basic concepts.  Vigorous scientific debate is often
beneficial in this way, though it becomes ineffective if criticisms are cited
while responses thereto are ignored.  A lack of clarity on the part of the
advocates of the histories approach during its first ten or fifteen years
contributed to the misunderstanding, but by now these earlier problems have
been cleared up.  It is hoped that the present paper, supplementing the
detailed exposition found in CQT, may serve to further understanding of an
approach to quantum foundations that deserves careful attention.  Along with
the measurement problem(s) it can resolve a host of quantum paradoxes: six
chapters are devoted to this in CQT. In addition it is consistent with special
relativity: there are no mysterious nonlocal influences.  No other approach to
quantum interpretation, at least none known to the author, can make comparable
claims.

\xb\outl{Quantum ontology: start with CM; what has changed?
}\xa

The approach to quantum ontology presented here starts by assuming that
classical mechanics, with its phase space and Hamiltonian equations of motion,
embodies much of what one might hope would be true of quantum mechanics: a
clean mathematical structure, an intuitive but reasonably plausible way to
associate the mathematics with (what realists believe to be) the ``reality out
there,'' and a system of interpretation in which human beings can seen as part
of, but not an essential component in, the physical world when described in
physical terms.  Of course, quantum mechanics must be different from classical
mechanics in some important way, as otherwise we quantum physicists have been
wasting our time. But whatever differences there are at the microscopic level,
the older classical ontology should be seen to emerge from, or at least be
consistent with, the more fundamental quantum perspective.

\xb\outl{Focusing on Cl to Qm changes avoids extraneous philosophical issues 
}\xa

Focusing on the changes needed when moving from the classical to the quantum
world has two fundamental advantages.  First, classical mechanics has been
around for a long time, and we can claim to understand it, and the associated
realistic ontology, reasonably well.  So our journey begins at a well-defined
location, rather than with complete ignorance.  Second, this route avoids
getting entangled in various philosophical issues, such as the ultimate
(un)reliability of human knowledge, which beset both classical and
quantum ontology. Putting them aside will allow a focus on a few
central issues, and the author to stay within areas where he can claim some
competence.

\xb\outl{The quantum ontology presented here
}\xa

\xb\outl{1 Has no measurement problem
}\xa

\xb\outl{2 Is fully consistent with textbook QT
}\xa

\xb\outl{3 Is local; no conflict with special relativity
}\xa

\xb\outl{4 Classical physics, logic emerge as appropriate approximations to QM
}\xa

\xb\outl{5 independent physical reality: observers/consciousness not needed
}\xa

\xb\outl{6 Provides a satisfactory foundation for Qm Info
}\xa

The quantum ontology presented below has the following features and
consequences.  First, it has no measurement problem; equivalently, it resolves
both measurement problems. Second, the results are fully consistent with
textbook formulations of quantum theory, once one comes to see that the
textbook approach provides a set of very successful and reasonably efficient
calculational tools, rather than a basic conceptual understanding of the
quantum world. It is now possible to see how these calculational tools arise
out of a fully consistent quantum perspective.  Third, quantum mechanics is a
\emph{local} theory in which mysterious nonlocal influences no longer play a
role \cite{Grff11}.  Thus quantum mechanics is fully consistent with special
relativity, as shown in \cite{Grff02b}, contrary to claims made in some
quarters. Fourth, the entire world of classical physics emerges, is consistent
with, quantum physics: classical mechanics in appropriate circumstances is an
approximation, sometimes an excellent approximation, to the underlying and
more exact quantum mechanics which encompasses all mechanical processes at
whatever length scale.  In circumstances where classical mechanics applies,
ordinary logic suffices for discussing physics, and can be seen to be
consistent with the more general mode of reasoning required in the quantum
domain.  Fifth, quantum mechanics is compatible with the traditional idea of
an \emph{independent} physical reality whose fundamental properties can be
discussed without needing to make reference to human observers or human
consciousness.  Sixth, this ontology provides a foundation for quantum
information; it supplies a specific answer to Bell's question as to what
quantum information is about.

\xb\outl{Price: unfamiliar form of reasoning; give up determinism
}\xa

All of this at what price?  First, the quantum world must be understood using
an appropriate form of reasoning with features which differ not only from
ordinary propositional logic, but also from the quantum logic proposed by
Birkhoff and von Neumann.  Second, determinism must be abandoned: quantum time
development is irreducibly stochastic in all circumstances, not just when
measurements occur.

\xb\outl{Organization of this paper. 
}\xa

\xb\outl{Single time in Sec. 2
}\xa

The remainder of the paper is organized as follows.  The ontology of a quantum
system at a single time is the topic of Sec.~\ref{sct2}, where it is developed
in analogy with classical phase space.  In particular, the ontology of Hilbert
space quantum mechanics necessarily differs from classical mechanics if one
follows (at least part way) von Neumann's interpretation of the basic quantum
formalism.  The logical problem this poses is discussed, along with the
solution proposed by the author and Omn\`es.

\xb\outl{Time development in Sec. 3
}\xa

Quantum time development is the topic of Sec.~\ref{sct3}.  Here the
fundamental idea goes back to Born\cite{Brn26}, but is further developed: the
proper use of Schr\"odinger's time dependent equation, unitary time
development, is to compute probabilities. The notion of a family of histories,
needed for a proper probabilistic framework of quantum dynamics, is
introduced, along with the technique needed for assigning probabilities in a
consistent way to histories inside a closed quantum system.

\xb\outl{Classical world, Sec. 4
}\xa

\xb\outl{Measurements, Sec. 5
}\xa

\xb\outl{Locality, Sec. 6
}\xa

\xb\outl{Ontology as a foundation of quantum information, Sec. 7
}\xa

Following Secs.~\ref{sct2} and \ref{sct3}, which form the heart of the paper,
some additional topics are discussed in a more cursory manner.  Any viable
quantum ontology must be able to make sense of the everyday ``classical''
world of our ordinary experience, and the strategy used to do this in the
histories approach is described in Sec.~\ref{sct4}. That program is not yet
complete, but nothing known at present seems to stand in the way of its full
realization, once a misunderstanding of the histories approach going back to
Dowker and Kent \cite{DwKn96} has been disposed of. Preparations and
measurements are the subject of Sec.~\ref{sct5}, which indicates the
essentials needed to resolve both measurement problems. Quantum locality,
including the validity of what is often referred to as Einstein locality, is
treated briefly in Sec.~\ref{sct6} (details will be found in
\cite{Grff11,Grff11b}).  Section~\ref{sct7} indicates in broad strokes how the
ontology presented here provides a foundation for quantum information.

\xb\outl{Conclusion in Sec. 8 
}\xa

\xb\outl{1. summary of logical issues
}\xa

\xb\outl{2. summary of indeterministic dynamics
}\xa

\xb\outl{3. Open issues
}\xa

Following a brief overall summary in Sec.~\ref{sbct8.1}, Sec.~\ref{sbct8.2} of
the concluding Sec.~\ref{sct8} is devoted to a discussion of the logical
issues which seem to be at the center of most criticisms of the histories
approach, and which need to be clearly understood, whatever conclusion the
reader may eventually wish to draw. Section~\ref{sbct8.3} contains a few
additional remarks about probabilistic dynamics. Finally some open issues, two
referring to the histories approach itself and two to its wider applications
to problems in the philosophy of science, are mentioned briefly in
Sec.~\ref{sbct8.4}

\xb
\section{System at One Time}
\label{sct2}
\xa

\begin{figure}[h]

$$
\begin{pspicture}(-7,-4)(7,3.5) 
\newpsobject{showgrid}{psgrid}{subgriddiv=1,griddots=10,gridlabels=6pt}
\def\lwd{0.035} 
\def\lwb{0.06}  
\def\rdot{0.13} \def\rodot{0.2} 
\psset{
labelsep=2.0,
arrowsize=0.150 1,linewidth=\lwd}
\def\dot{\pscircle*(0,0){\rdot}} 
\def\odot{\pscircle[fillcolor=white,fillstyle=solid](0,0){\rodot}} 
		\def\pspace{
\psline{->}(-3,0)(3,0)
\psline{->}(0,-3)(0,3)
\rput[l](3.1,0){$x$}
\rput[b](0,3.1){$p$}
\pscircle[linewidth=\lwb](-1.8,-0.1){1.5}
\pscircle[linewidth=\lwb](-0.8,-1.5){1.2}
\rput[t](-1.8,1.1){$\PM$}
\rput[b](-0.8,-2.4){$\QM$}
\rput[B](0,-3.7){(a)}
		}
		\def\hspace{
\psline{->}(-3,0)(3,0)
\psline{->}(0,-3)(0,3)
\psline[linewidth=\lwb](-3,-1)(3,1)
\psline[linewidth=\lwb](-1,3)(1,-3)
\psline[linewidth=\lwb,linestyle=dashed](-3,1)(3,-1)
\rput[b](2,0.9){$\PM$}
\rput[r](-0.9,2){$\PM^\perp$}
\rput[t](2,-1){$\QM$}
\rput[B](0,-3.7){(b)}
		}
\rput(3.5,0){\hspace}
\rput(-3.5,0){\pspace}
\end{pspicture}
$$
\caption{%
(a) Classical phase space; (b) Hilbert space}
\label{fgr1}
\end{figure}
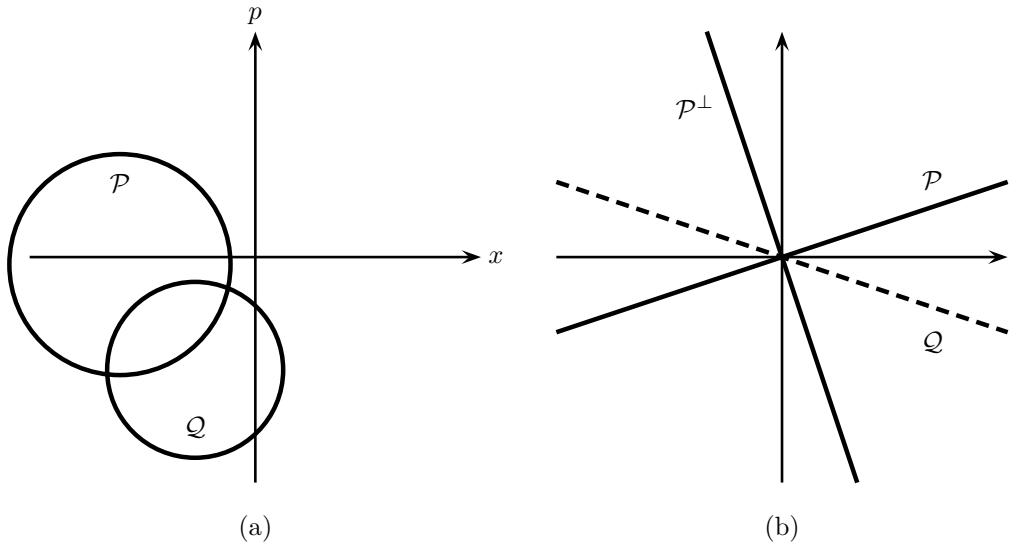

\xb
\subsection{Phase Space and Hilbert Space}
\label{sbct2.1}
\xa

\xb\outl{Classical phase space as a model
}\xa

Starting with classical phase space and classical dynamics, what changes are
needed to arrive at the corresponding concepts for quantum theory?  The phase
space (position $x$, momentum $p$) of a particle moving in one dimension is
shown in Fig.~\ref{fgr1}(a), while (b) is a somewhat schematic representation
of a two-dimensional (complex) Hilbert space,%
\footnote{We follow the convention of modern quantum information theory, where
  the term ``Hilbert space'' is not restricted to infinite-dimensional spaces,
  but also includes any finite-dimensional complex linear vector space
  equipped with an inner product. By restricting the following discussion to
  finite-dimensional Hilbert spaces we avoid various technical issues that are
  not pertinent to the conceptual issues we are concerned with.} %
the closest quantum counterpart to phase space, for the simplest of quantum
systems: the spin angular momentum of a spin-half particle (a single qubit in
the jargon of quantum information theory).

\xb\outl{Cl prop. $\lra$ subset, indicator; Qm prop. $\lra$ ray, sspace, projector
}\xa

Following von Neumann \cite{vNmn32b} we assume that a single point in
classical phase space corresponds to a ray or one-dimensional subspace in the
Hilbert space: all multiples of a nonzero $\ket{\psi}$ by an arbitrary
(complex) number.  A ray is a single line through the origin in part (b) of
the figure. In addition, the counterpart of a \emph{classical property} $P$,
such as ``the energy is between 1.9 and 2.0 Joules'', represented by some
(measurable) collection of points $\PM$ in the classical phase space, is a
\emph{quantum property} represented by a (closed) \emph{subspace} of the
quantum Hilbert space.  The classical property can be represented by an
\emph{indicator} function $P(\gm)$, where $\gm$ denotes a point in the
classical phase space, and $P(\gm)$ is 1 at all points where the property
holds or is true, and 0 at all other points.  The quantum counterpart of an
indicator function is a \emph{projector}, a Hermitian operator on the Hilbert
space that is equal to its square, so its eigenvalues are 0 and 1. It projects
onto the subspace corresponding to the quantum property; any ket in this
subspace is an eigenvector with eigenvalue 1.  The negation $\lnot P$, \NOT\
$P$, of a classical property corresponds to the set-theoretic complement
$\PM^c$ of $\PM$, with indicator $I-P$; $I$ is the function whose value is 1
everywhere on the phase space. Following von Neumann we assume that the
negation of a quantum property is the \emph{orthogonal complement} $\PM^\perp$
of the corresponding subspace $\PM$, with projector $I-P$, where $I$ is the
identity operator on the Hilbert space.

\xb\outl{Reasons for accepting vN proposal of complementary sspace for negation
}\xa

\xb\outl{Accepted by physicists, consistent with what is taught to students
}\xa

\xb\outl{Mathematically natural: inner product $\lra$ orthogonal
}\xa

\xb\outl{Orthogonal complement,  not set-theoretical complement, of a subspace
  is a subspace
}\xa

Von Neumann's proposal for associating the negation of a quantum property with
the complementary subspace is central to our construction of a consistent
quantum ontology, and hence it is appropriate to discuss why this approach is
reasonable and to be preferred to the alternative of letting the negation of a
property consist of the set-theoretic complement of the ray or subspace.
First, the von Neumann approach has been widely accepted and is not (so far as
the author is aware) contested by physicists, though it is not always accepted
in the quantum foundations community. (We are dealing with an example of the
eigenvalue-eigenvector link.) It is consistent with the way students are
taught in textbooks to make calculations, even though the idea itself is
(alas) not always included in textbook discussions.  Second, it is
mathematically ``natural'' in that it makes use of a central property, the
inner product, of the Hilbert space, which is what distinguishes such a space
from an ''ordinary'' complex linear vector space.  (The inner product defines
what one means by ``orthogonal.'')  Third, the orthogonal complement of a
subspace is a subspace, whereas the set-theoretic complement is not a
subspace; see, e.g., Fig.~\ref{fgr1}(b).

\xb\outl{Spin half. vN proposal consistent with SG expt, with chemistry
}\xa

Fourth, in the case of a spin-half particle the von Neumann proposal asserts
that the complement or negation of the property $S_z=+1/2$ (in units of
$\hbar$) is $S_z = -1/2$, which is to say one of these properties is the
negation of the other, in accord with the Stern-Gerlach experimental result,
which showed, somewhat to the surprise of the physics community at the time,
that silver atoms passing through a magnetic field gradient come out in
\emph{two} distinct beams, not the infinite number which would have been
expected for a classical spinning particle. Or might still be expected for a
spin-half quantum particle, were one to assume that all the other rays in the
two-dimensional Hilbert space represent alternative possibilities to
$S_z=+1/2$.  However, the Stern-Gerlach result is in complete accord with von
Neumann's approach: the negation of the property corresponding to a particular
ray in a two-dimensional Hilbert space is the unique property defined by the
orthogonal ray, and a measurement determines which of these two properties is
correct in a given case. 

\xb\outl{Chemistry and statistical mechanics: spin half particle has two states
}\xa

Fifth there is a sense in which all of chemistry is based on the idea that the
electron, a spin-half particle, has only two possible spin states: one ``up''
and one ``down.''  Granted, students of chemistry find this confusing, since
it is not clear how ``up'' and ``down'' are to be defined, though they
eventually are bludgeoned into shutting up and calculating.  Among the
calculations which lead to the wrong answer if 2 is replaced by some other
number is that of finding the entropy of a partially ionized gas, where it is
important to take into account all degrees of freedom, and log 2 is the
correct contribution from the electron spin.  In addition, the modern theory
of quantum information is consistent with the idea that a single qubit with
its two-dimensional Hilbert space can contain at most one bit of information;
see \cite{Grff02}.

\xb\outl{How shall we think about rays neither in sspace or its orthogonal complement?
}\xa

It is important to stress this point, for if one accepts von Neumann's
negation a wide gap opens between classical phase space and quantum Hilbert
space, one clearly visible in Fig.~\ref{fgr1}.  Any point in classical phase
space, any possible mechanical state of the system, is either inside or
outside the set that defines a physical property, which is therefore either
true or false.  In the quantum case there are vast numbers of rays in the
Hilbert space that lie neither inside the ray or, more generally in higher
dimensions, the subspace corresponding to a property, nor in the complementary
subspace that corresponds to its negation. For these the property seems to be
neither true nor false, but somehow undefined.  How are we to think about this
situation?  This is in a sense \emph{the central issue} for any ontology that
uses the \emph{quantum Hilbert space} as the physicist's fundamental
mathematical tool for representing the quantum world.  One approach, typical
of textbooks, is to ignore the problem and instead take refuge in
``measurements.''  But problems do not go away by simply being ignored, and
ignoring this particular problem makes it re-emerge under a different name:
the measurement problem.

\xb\outl{Issue restated using projectors which in Qm case may not commute
}\xa

Before going further, let us restate the matter in the language of indicator
functions and projectors introduced earlier.  The product of two (classical)
indicator functions $P$ and $Q$ is itself an indicator function for the
property $P$ \AND\ $Q$, the intersection of the two sets of points $\PM$ and
$\QM$ in the phase spaces, Fig.~\ref{fgr1}(a). However, the product of two
(quantum) projectors $P$ and $Q$ is itself a projector \emph{if and only if
  $PQ=QP$}, that is if the projectors \emph{commute}.  Otherwise, when $PQ$ is
not equal to $QP$, neither product is a projector, and thus neither can
correspond to a quantum property. A ray $\QM$ in a two-dimensional Hilbert
space, Fig.~\ref{fgr1}(b), that coincides neither with a ray $\PM$ nor its
orthogonal complement $\PM^\perp$ is represented by a projector $Q$ that does
not commute with $P$, and thus it is unclear how to define the conjunction of
the corresponding properties.

\xb\outl{Birkhoff and von Neumann approach
}\xa

Von Neumann was not unaware of this problem, and he and Birkhoff had an idea
of how to deal with it.  Instead of using the product of two noncommuting
projectors it seemed to them sensible to employ the set theoretic
\emph{intersection} of the corresponding subspaces of the Hilbert space, which
is itself a subspace and thus corresponds to some property, as a sort of
quantum counterpart of \AND.  The corresponding \OR\ can then be associated
with the direct sum of the two subspaces.  When projectors for the two
subspaces commute these geometrical constructions lead to spaces whose quantum
projectors, $PQ$ and $P+Q-PQ$, coincide precisely with what one would expect
based on the analogy with classical indicators.

\xb\outl{Quantum logic has not solved the conceptual problems. Robots might help
}\xa

The resulting structure, known as \emph{quantum logic}, obeys some but not all
of the rules of ordinary propositional logic, as Birkhoff and von Neumann
pointed out.  If one naively goes ahead and applies the usual rules of
reasoning with these definitions of \AND\ and \OR\ it is
easy to  construct a contradiction; for a simple example see Sec.~4.6 of
CQT.  While quantum logic has been fairly extensively studied it seems fair to
say that this route has not made much if any progress in resolving the
conceptual difficulties of quantum theory. It is not mentioned in most
textbooks, and is given negligible space in \cite{Wllc08}. Perhaps the
difficulty is that physicists are simply not smart enough, and the
resolution of quantum mysteries by this route will have to await the
construction of superintelligent robots. (But if the robots succeed, will
they be able to, or even interested in, explaining it to \emph{us}?)

\xb\outl{The CH approach: only accept conjunctions if PQ=QP
}\xa

The histories approach follows von Neumann in letting subspaces represent
properties, with the negation of a property represented by the orthogonal
complement of the subspace.  But it takes a very different and much more
conservative attitude than Birkhoff and von Neumann in the case of properties
$P$ and $Q$ represented by noncommuting projectors.  Since neither $PQ$ nor
$QP$ is a projector, let us not try and talk about their conjunction.  Let us
adopt a language for quantum theory in which $PQ$ makes sense and represents
the conjunction of the two properties in those cases in which $QP=PQ$.  But if
$PQ\neq QP$ the statement ``$P$ \AND\ $Q$'' is \emph{meaningless}, not in a
pejorative sense but in the precise sense that this interpretation of quantum
mechanics can assign it no meaning.  In much the same way that in ordinary
logic the proposition $P\land\lor\, Q$, ``$P$ \AND\ \OR\ $Q$,'' is meaningless
even if $P$ and $Q$ make sense: the combination $P\land\lor\, Q$ has not been
put together using the rules for constructing meaningful statements.
Likewise, in quantum mechanics both the conjunction and also the disjunction
$P$ \OR\ $Q$, must be excluded from meaningful discussion if the projectors do
not commute.

\xb\outl{Meaningless vs false.  Example of $S_x =\cdots$  AND $S_z =\cdots$
}\xa

The two-dimensional Hilbert space of a spin-half particle can serve as an
illustration.  Let
\begin{equation}
\label{eqn1}
 [z^+] := \dya{z^+},\quad [z^-] := \dya{z^-},
\end{equation}
where we shall (as in CQT) hereafter employ the $[\cdot]$ notation for this
type of dyad, be projectors for the properties $S_z = +1/2$ and $S_z = -1/2$;
similarly $[x^+]$ and $[x^-]$ are the corresponding properties for $S_x$.  The
product of $[z^+]$ and $[z^-]$ is zero (i.e., the zero operator) in either
order, so they commute and their conjunction is meaningful: it is the quantum
property that is always false (the counterpart of the empty set in the case of
a phase space).  However, neither $[z^+]$ nor $[z^-]$ commutes with either
$[x^+]$ or $[x^-]$, so the conjunction ``$S_x = +1/2$ AND $S_z = +1/2$'' is
meaningless.  In support of the claim that it lacks meaning one can note that
every ray in the spin-half Hilbert space has the interpretation $S_w=+1/2$ for
some direction in space $w$.  Thus there are no spare rays available to
represent ``$S_x = +1/2$ \AND\ $S_z = +1/2$''; there is no room in the Hilbert
space for such conjunctions.  It is very important to distinguish
``meaningless'' from ``false''.  In ordinary logic if a statement is false
then its negation is true.  But the negation of a meaningless statement is
equally meaningless.  Thus the statement ``$S_z=+1/2$ \AND\ $S_z=-1/2$'' is
meaningful and always false, whence its negation ``$S_z=-1/2$ \OR\
$S_z=+1/2$'' is meaningful and always true, and is consistent with the
Stern-Gerlach experiment.  On the other hand, ``$S_x=+1/2$ \AND\ $S_z=-1/2$''
is meaningless, and its formal negation, ``$S_x=-1/2$ \OR\ $S_z=+1/2$,'' is
equally meaningless.

\xb\outl{$S_x =\cdots$  OR $S_z =\cdots$ also meaningless
}\xa

The student who has learned quantum theory from the usual courses and
textbooks may well go along with the idea that ``$S_x = +1/2$ \AND\ $S_z =
+1/2$'' lacks meaning, since he can think of no way of measuring it (and has
probably been told that it cannot be measured).  However he will be less
likely to go along with the equally important idea that the disjunction ``$S_x
= +1/2$ \OR\ $S_z = +1/2$'' is similarly meaningless.  Granted, there is no
measurement which can distinguish the two, but he has a mental image of a
spin-half particle in the state $S_z=+1/2$ as a little gyroscope with its axis
of spin coinciding with the $z$ axis.  Such mental images are very useful to
the physicist, and perhaps indispensable, for they help organize our picture
of the world in terms that are easily remembered; they provide ``physical
intuition.''  But this particular mental image can be quite misleading in
suggesting that when $S_z=+1/2$ the orthogonal components of angular momentum
are zero.  However, this cannot be the case since, since as the student has
been taught, $S_x^2 = S_y^2 = 1/4$.  Now any classical picture is bound to
mislead to some extent when one is trying to think about the quantum world,
but in this case a slight modification is less misleading.  Imagine a
gyroscope whose axis is oriented at random, except one knows that the $z$
component of angular momentum is positive rather than being negative. Among
other things this modified image helps guard against the error that the spin
degree of freedom of a spin-half particle can carry a large amount of
information, when in fact the limit is one bit ($\log_2 2$).

\xb
\subsection{Frameworks}
\label{sbct2.2}
\xa

\xb\outl{Probability theory uses sample space. Coarse graining of Cl phase space
}\xa

\xb\outl{Classical indicator functions
}\xa

\xb\outl{Cl stat mech employs Borel sets, etc.; we do not need these complications
}\xa

Ordinary probability theory uses the concept of a \emph{sample space}: a
collection of mutually exclusive alternatives or events, one and only one of
which occurs or is true in a particular realization of some process or
experiment. One way of introducing probabilities in classical statistical
mechanics is to imagine the phase space divided up into a collection of
nonoverlapping cells, a \emph{coarse graining} in which each cell represents
one of the mutually exclusive alternatives one has in mind.  Let $P^j(\gm)$ be
the indicator function for the $j$'th cell: equal to 1 if $\gm$ lies within
the cell and 0 otherwise.  (We are using the superscript of $P$ for a label,
not an exponent; as the square of an indicator is equal to itself, this need
not cause confusion.) Obviously the product $P^j P^k$ of the indicator
functions for two different cells is 0, and the sum of all the indicator
functions is the identity function $I(\gm)$ equal to 1 for all $\gm$:
\begin{equation}
   P^j P^k = \dl_{jk} P^j;\quad \sum_j P^j = I.
\label{eqn2}
\end{equation}

\xb\outl{Boolean event algebra of Cl events; probability assignment
}\xa

Next, probabilities are assigned to the events making up a Boolean \emph{event
  algebra}.  If one coarse grains the phase space in the manner just
indicated, an event algebra can be constructed in which each event is
represented by the union of some of the cells in the coarse graining;
equivalently, the event algebra is the collection of all indicator functions
which are sums of some of the indicators in the collection $\{P^j\}$,
including the functions $\emptyset$ and $I$ which are everywhere 0 and 1,
respectively.  The probabilities themselves can be specified by a collection
$\{p_j\}$ of nonnegative real numbers that sum to 1, with the probability of
an event $E$ being the sum of those $p_j$ for which the corresponding
indicator functions $P^j$ appear in the sum defining $E$.  To be sure,
classical statistical mechanics is usually constructed without using a coarse
graining, employing the Borel sets as an event algebra, and then introducing
an additive positive measure to define probabilities.  There is nothing wrong
with this, but for our purposes a coarse graining provides a more useful
classical analog.

\xb\outl{Quantum PD or framework is counterpart of Cl sample space
}\xa

\xb\outl{Boolean event algebra formed from PD is also a 'framework'
}\xa

The quantum counterpart of a classical sample space is referred to in the
histories approach as a \emph{framework}.  It is a \emph{projective
  decomposition} (PD) of the identity operator I: a collection $\{P^j\}$ of
mutually orthogonal projectors which sum to the identity operator $I$, and
thus formally satisfy exactly the same conditions \eqref{eqn2} as a collection
of classical indicators.  The fact that $P^j P^k$ vanishes for $j \neq k$
means that the corresponding quantum properties (``events'' is the customary
term in probability theory) are mutually exclusive: if one is true the other
must be false, and the fact that they sum to the identity operator $I$ means
that at least one, and therefore only one, is true or real or actual.  The
corresponding event algebra is the Boolean event algebra of all projectors
which can be formed by taking sums of projectors in $\{P^j\}$ along with the 0
operator, which plays the same role as the empty set in ordinary probability
theory.  As long as the quantum event algebra and the sample space are related
in this way, there is no harm in using the somewhat loose term ``framework''
to refer to either one, as we shall do in what follows.

\xb\outl{Compatible and incompatible frameworks. Refinements
}\xa

\xb\outl{Two refinements of single framework can be incompatible 
}\xa

\xb\outl{Importance of keeping track of Qm framework
}\xa

Two frameworks $\{P^j\}$ and $\{Q^k\}$ are \emph{compatible} if all the
projectors in one commute with all of those in the other: $P^j Q^k = Q^k P^j$
for all values of $j$ and $k$. Otherwise they are \emph{incompatible}.  One
says that $\{P^j\}$ is a refinement $\{Q^k\}$ if the projectors in the PD of
the latter are included in the event algebra of the former; equivalently,
$\{Q^k\}$ is a coarsening of $\{P^j\}$. Obviously two frameworks must be
compatible if one is to be a refining or coarsening of the other. In addition,
two compatible frameworks always possess a \emph{common refinement} using the
PD consisting of all the nonzero products $P^j Q^k$; its event algebra
includes the union of the event algebras of the separate frameworks which it
refines.
Note that a given framework $\{P^j\}$ may have various different refinements,
and two refinements need not be compatible with each other.  Therefore when
discussing quantum systems one must keep track of the framework being employed
in a particular argument.  This is not important in classical physics,
where one can either adopt the finest framework possible at the outset, or
else refine it as one goes along without needing to call attention to this
fact.  In quantum mechanics one does not have this freedom, and carelessness
can lead to paradoxes.

\xb
\subsection{The Single Framework Rule}
\label{sbct2.3}
\xa

\xb\outl{Single framework rule is central concept in histories approach
}\xa

A central concept of the histories approach is the \emph{single framework
  rule}, which states that probabilistic reasoning that starts from data
(observed or simply assumed) about a quantum system and leads to conclusions
about the same system, typically expressed as conditional probabilities, is
invalid unless it is carried out using a \emph{single framework} as defined
above. In particular it is not valid when it results from combining
\emph{incompatible} frameworks.

\xb\outl{Single framework rule has been badly misunderstood
}\xa

\xb\outl{Physicist at Liberty to create different descriptions
}\xa

\xb\outl{Equality of descriptions from a fundamental perspective
}\xa

\xb\outl{Incompatibility: only compatible frameworks can be combined
}\xa

\xb\outl{Utility: not all frameworks equally useful
}\xa

As the single framework rule has been frequently misunderstood by critics of
the histories approach, it is important to clarify what it does and does not
mean.  While it rules out improper \emph{combinations} of descriptions, it
does not prevent the physicist from employing a variety of different and
possibly incompatible frameworks when constructing several distinct
descriptions of a quantum system, each of which may provide some physical
insight about its behavior. This is the principle of Liberty: the physicist
can use whatever framework he chooses when describing a system.  All properly
constructed individual frameworks are equally acceptable in terms of
fundamental quantum mechanics: the principle of Equality.  The principle of
Incompatibility forbids combining incompatible frameworks.  Not all frameworks
are equally useful in answering particular questions of physical interest, let
us call this the principle of Utility.  It is by combining these principles
that the single framework rule arrives at a consistent quantum ontology
adequate for understanding the quantum world in a realistic way, while at the
same time resolving or avoiding (or ``taming'') the numerous paradoxes or
inconsistencies that beset alternative approaches to quantum interpretation.

\xb\outl{Liberty: Can discuss position or rate of rotation of Jupiter. 
}\xa

\xb\outl{$S_x$, $S_z$ for a silver atom: both possible, but cannot be combined.
}\xa

The physicist's Liberty to choose different frameworks should not be thought of
as in any way influencing reality.  Choosing a description is choosing what to
talk about, and what physicists choose to talk about has very little (direct)
influence on what actually goes on in the world.  Shall we discuss the
location of Jupiter's center of mass, or its rate of rotation?  Either is
possible, and neither has the slightest influence on the behavior of
Jupiter. What about a silver atom approaching a Stern-Gerlach apparatus?
Shall we discuss its (approximate) location or the value of $S_x$ or the value
of $S_z$?  Any one of these is possible, and the single framework rule allows
location to be combined with $S_x$, or with $S_z$.  But not with both, since
it is impossible to put both $S_x$ and $S_z$, at least when referring to a
single particle at a particular time, in the same framework.  In this case, no
less than for Jupiter, the physicist's choice of framework has not the
slightest influence on the silver atom. And there is no law of nature which
singles out a framework that includes $S_x$ as somehow ``correct'' or ``true''
in distinction to the $S_z$ framework; that would contradict Equality.
However, they are incompatible.  They cannot be combined.  What does this
mean?

\xb\outl{Incompatibility does not mean mutually exclusive, like two sides of a coin
}\xa

\xb\outl{Instead, combinations, rather than being false, are meaningless
}\xa

Incompatibility in this technical sense is a feature of the quantum world with
no exact classical analog: in classical physics all the operators commute.
But classical analogies and disanalogies, together with applications to
various specifically quantum situations can help tease out its intuitive
meaning.  Let us start with a disanalogy.  A coin can land heads or tails, two
mutually exclusive possibilities: if one is true, the other must be
false. There is a temptation to think of the relationship of the incompatible
$S_x$ and $S_z$ frameworks in this way, and it must be resisted, for it leads
to a serious misunderstanding.  The properties $S_z=+1/2$ and $S_z=-1/2$ are
analogous to heads and tails: they are mutually exclusive.  If one is true the
other is false, and the combination ``$S_z=+1/2$ \AND\ $S_z=-1/2$'' is
meaningful and false, as discussed earlier.  On the other hand the combination
``$S_x=+1/2$ \AND\ $S_z=-1/2$'' is meaningless, neither true nor false.
Statements belonging to incompatible frameworks cannot be compared in any way,
which is also why it is meaningless to say that one framework rather than the
other is the true or correct way of describing the quantum world.  Equality
must be taken seriously.

\xb\outl{Classical coarse graining as an analogy
}\xa

For a positive analogy, think of a framework as something like a coarse
graining of the classical phase space as discussed earlier in
Sec.~\ref{sbct2.2}.  Many coarse grainings are possible and the physicist is
at Liberty to choose one that is convenient for whatever purposes he has in
mind.  There is no ``correct'' coarse graining, though some coarse grainings
may be more useful than others in discussing a particular problem.  The
physicist's choice of coarse graining does not, of course, have any influence
on the system whose properties he is trying to model.  In all these respects
the choice of coarse graining is like the choice of a quantum framework.  But
classical coarse grainings of the same phase space can always be combined: the
common refinement is constructed in an obvious way using cells formed by
intersections of those taken from the two coarse grainings that are being
refined.  However, PDs of the same Hilbert space cannot in general be
combined, so in this respect the analogy fails.  However, it is still helpful
in illustrating some aspects of the quantum situation, and in avoiding the
misleading idea that the relationship between different quantum frameworks is
one of mutual exclusivity.

\xb\outl{Analogy of different inertial frames in special relativity
}\xa

Similarly, choosing a framework is something like choosing an inertial
reference frame in special relativity.  The choice is up to the physicist,
and there is no law of nature, at least no law belonging to relativity theory,
that singles out one rather than another.  Sometimes one choice is more
convenient than another when discussing a particular problem; e.g., the
reference frame in which the center of mass is at rest.  The choice obviously
does not have any influence upon the real world.  But again there is a
disanalogy: any argument worked out using one inertial frame can be worked out
in another; the two descriptions can be mapped onto each other.  This is not
true for quantum frameworks: one must employ a framework (there may be several
possibilities) in which the properties of interest can be described; they must
lie in the event algebra of the corresponding PD.

\xb\outl{Different views of Mt Rainier. Unicity
}\xa

For a more picturesque positive analogy consider a mountain, say Mount Rainier,
which can be viewed from different sides.  An observer can choose to look at it
from the north or from the south; there is no ``law of nature'' that singles
out one perspective as the correct one.  One can learn different things from
different viewpoints, so there might be some Utility in adopting one
perspective rather than the other.  But once again the analogy fails in that
the north and south views can, at least in principle, be combined into a single
unified description of Mount Rainier from which both views can be derived as
partial descriptions.  Let us call this the \emph{principle of unicity}.  It no
longer holds in the quantum world once one assumes the Hilbert space represents
properties in the manner discussed above.

\xb\outl{Utility: Silver atom between preparation and measurement
}\xa

But how can Utility play a role in quantum physics?  Again consider the case
of a spin-half silver atom, and suppose it is midway in its trajectory from an
apparatus where a competent experimentalist has prepared it in a state with
$S_x=+1/2$ to an apparatus, also constructed by a competent experimentalist,
which will later measure $S_z$ with the pointer corresponding to $S_z = +1/2$.
What can one say about the spin of the atom midway between preparation and
measurement, assuming it travels in a region free from magnetic fields that
could cause the spin to precess?  There is a framework which at the
intermediate time includes the possibilities $S_x = +1/2$ and $-1/2$; using
this $S_x$ framework and the data about the preparation one can infer that
$S_x = +1/2$ with probability 1 and $S_x = -1/2$ with probability 0. There is
an alternative $S_z$ framework that at the intermediate time includes the
possibilities $S_z = +1/2$ and $-1/2$, and it can be used to infer from the
later measurement outcome that at the intermediate time the (conditional)
probabilities for $S_z = +1/2$ and $-1/2$ are 1 and 0 respectively.  The $S_x$
framework is useful if one is concerned with whether the preparation
apparatus was functioning properly, while the $S_z$ framework is useful if one
wants to discuss the proper functioning of the measuring device.

\xb
\section{Time Development}
\label{sct3}
\xa

\xb
\subsection{Quantum dynamics: histories}
\label{sbct3.1}
\xa

\xb\outl{histories approach uses stochastic time development
}\xa

\xb\outl{example of decay of atoms and nuclei: no clock sets the decay time
}\xa

\xb\outl{Bohmian mechanics is exception
}\xa

The histories approach treats the time dependence of a quantum system as a
random or stochastic process, one in which the future and past states of the
system at different times are not determined by the present state, but only
related to it by certain probabilities, which only in very special cases are 0
and 1, corresponding to a deterministic time development.  Indeed, most
physicists accept that in practice quantum-mechanical time development is
probabilistic and not deterministic.  Take the case of spontaneous decay.
Modern physics possesses tools for calculating decay rates of atoms in excited
states, or unstable nuclei, and they work reasonably well.  If the quantum
world were deterministic one would expect to find somewhere in the theoretical
formalism a prescription for predicting the (relatively) precise time of decay
of, say a radioactive nucleus.  This time can be measured quite precisely,
better than a millisecond, for a nucleus with a half life of minutes or hours
or even years.  Most quantum physicists do not believe that there is some
``marker'' or ``clock'' inside the nucleus which before the decay can be used
to the time of decay; there is no room for it in the Hilbert space
description.%
\footnote{A notable exception is the proponents of Bohmian mechanics, who
  add additional ``hidden'' variables to the Hilbert space description.  See,
  e.g., \cite{Glds06}. However, in the case of radioactive decay they have no
  way of accessing this deterministic internal clock other than by observing
  the actual time of decay. Determinism is also upheld by followers of
  Everett, see, e.g., \cite{Brrt09,Vdmn09}, but they, too, have no way of
  accessing the alternative worlds (or minds or whatever) in which the nucleus
  decays at a time different from that observed in the laboratory.} %
Hence the assumption of a probabilistic dynamics does not, in the modern
context, represent much of an innovation.  But doing it in a consistent way
that avoids paradoxes is not altogether straightforward.

\xb\outl{Sample space, event algebra discussed earlier in Sec. 2
}\xa

\xb\outl{There is no general rule for assigning probabilities
}\xa

The essential mathematical structures needed for introducing probabilities
into quantum mechanics were introduced in Sec.~\ref{sbct2.2}: a sample space
constituted by a projective decomposition (PD) of the identity operator I on
the Hilbert space, and a Boolean event algebra of projectors generated from
the PD in a natural way.  What remains is to assign nonnegative probabilities
$p_j$ summing to 1 to the elements $P^j$ of the sample space, and thereby to
the subspaces (events) that constitute the event algebra in precisely the same
way as in other applications of probability theory.  If this can be done, one
can then carry out probabilistic reasoning in the quantum domain following all
of the ordinary rules of probability theory \emph{provided the sample space
  (and event algebra) remain fixed while various (conditional) probabilities
  are computed}---the single framework rule.  Just as in other applications of
probability theory there is no general rule that specifies the probabilities
$p_j$: they enter a probabilistic description of the world as parameters, and
some exercise of judgment on the part of the scientist is generally necessary,
as well as input data, results of experiments, etc.

\xb\outl{QM has special rule for time development of closed system
}\xa

However, quantum theory introduces a new element not found in other
applications of probability theory.  It applies to the time development of a
\emph{closed} quantum system, by which we mean either that it is isolated,
completely self-contained with no environment with which it interacts, or that
its interaction with with its environment is well-enough approximated by
assigning to the system itself a (possibly time-dependent) Hamiltonian, which
can then be employed in Schr\"odinger's equation.  In this case certain
(conditional) probabilities relating states of the system at different times
can be assigned through the Born rule and its extensions to more than two
times, as discussed in Secs.~\ref{sbct3.3} and \ref{sbct3.4}.  The unitary
time evolution induced by Schr\"odinger's equation is used in calculating
these probabilities, but only in exceptional cases, as discussed in
Sec.~\ref{sbct3.2}, can it be used directly (without a probabilistic
interpretation) to describe what is really going on in a closed quantum
system.

\xb\outl{Sample space of histories.  Classical: several times = several systems
}\xa

Before assigning probabilities to processes occurring in time, we
need to construct an appropriate quantum sample space.  How is this to be
done?  Classical physics provides a useful hint.  The sample space
for a coin tossed three times in a row consists of the eight possibilities
$HHH$, $HHT$,\dots\ $TTT$, where $HTT$ means heads on the first toss, tails on
the second and third.  This is just the Cartesian product of the sample space
for a single toss of the coin, and is in fact identical to the sample space
needed to describe three different coins all tossed at the same time.  Hence
in ordinary probability theory sequences of events at successive times
in a particular system are formally the same thing as multiple copies of
the same system considered at a single time.

\xb\outl{Tensor product space of histories: spin-half example
}\xa

In quantum mechanics the mathematics for describing a compound system
consisting of a collection of (distinguishable) subsystems is well know: one
uses the tensor product of the Hilbert spaces.  This immediately suggests that
the way to construct a quantum sample space for a system at successive times
is to use a tensor product of copies of its Hilbert space, as first proposed
in \cite{Ishm94}.  Thus for a spin-half particle the 8-dimensional Hilbert
space
\begin{equation}
\breve \HM = \HM_0 \odot \HM_1 \odot \HM_2 
\label{eqn3}
\end{equation}
is appropriate for describing its properties at three successive times, $t_1 <
t_2 < t_3$, where each $\HM_m$ is a copy of the two-dimensional Hilbert space
needed to describe it at one time, and following CQT we use the symbol $\odot$
in place of the customary $\otimes$ symbol---formally they mean the same
thing---in order to emphasize that we are considering a sequence of successive
times, rather than several systems at the same time.  A spin-half particle
that possesses property $F_m$ at time $t_m$ is then described by a
\emph{history projector}
\begin{equation}
  F_0 \odot F_1 \odot F_2,
\label{eqn4}
\end{equation}
where each $F_m$ is a projector on the two-dimensional single-time Hilbert
space.  One might, for example, suppose that each $F_m$ is $[z^+]$ or $[z^-]$.
This is rather like the case of flipping a coin three times: think of $[z^+]$,
meaning $S_z = +1/2$ as heads, $[z^-]$ as tails.  There are eight
possibilities, and summing the eight projectors indeed yields $\breve I$, the
identity operator on $\breve \HM$.  Each element in the sample space is called
a \emph{history}, and together they constitute a \emph{family of
  histories}. Let us call this family $\FM_1$. An alternative family can be
constructed using the alternatives $[z^+]$ and $[z^-]$ at $t_0$, but then
the alternatives $[x^+]$ and $[x^-]$ at times $t_1$ and $t_2$.  This second
family $\FM_2$ corresponds to a different PD of $\breve I$, and $\FM_2$ is
incompatible with $\FM_1$ because  the two sets of projectors,
regarded as operators on $\breve \HM$,  do not commute with each other.

\xb\outl{Histories $Y^\al$ for an arbitrary collection of times
}\xa

The generalization to an arbitrary but finite collection of $f+1$ times $t_0 <
t_1 < \cdots t_f$ proceeds in an obvious way. First define the \emph{histories
  Hilbert space}, the obvious generalization of \eqref{eqn3} to $f+1$ copies
of the single-time Hilbert space. Next introduce a \emph{sample space}
$\{Y^\al\}$ of histories of the form
\begin{equation}
  Y^\al = F_0^{\al}\od F_1^{\al}\od \cdots  F_f^{\al},
\label{eqn5}
\end{equation}
where the superscript $\al$ labels the different elements of the sample space,
$F_m^\al$ is a projector representing the a property of the quantum system at
time $t_m$, and the histories projectors satisfy the condition
\begin{equation}
 \sum_\al Y^\al = \breve I,
\label{eqn6}
\end{equation}
the counterpart of \eqref{eqn2}, where $\breve I = I_1\ot I_2\ot\cdots$ is the
identity operator on the histories Hilbert space $\breve\HM$.  This ensures
that $Y^\al Y^\bt=0$ if $\al\neq \bt$, i.e., two distinct histories belonging
to this sample space are mutually exclusive.

\xb\outl{Special cases: fixed initial state, fixed PD at each time
}\xa

Two special cases are worth mentioning.  The first is that of a \emph{fixed
  initial state}: $F_0^\al$ at time $t_0$ is equal to the same projector, call
it $P_0$, for all $\al$ with the single exception of a particular history
labeled $\al=0$. For the $\al=0$ history, $F_0^0=(I-P_0)$ at time $t_0$, and
$F_m^0=I$ at all later times $t_m> t_0$.  This special history, whose physical
significance is that $P_0$ is \emph{not} true at $t_0$, is a throwaway: one
sets its probability to zero; it has been introduced simply to ensure that
\eqref{eqn6} is satisfied.  One then fixes attention on cases in which the
system is initially in the state $P_0$, which may be a pure state but may also
correspond to a subspace of dimension greater than one.  A second special case
is the one in which at each time $t_m$ the projector $F_m^\al$ is drawn from a
fixed PD $\{P_m^j\}$ which can be different for different times $t_m$.
Although these special cases, often combined, receive the most attention in
CQT, there are also other interesting examples.

\xb\outl{Many possible frameworks or families of histories
}\xa

Thus the situation for describing a quantum system at multiple times is
closely analogous to describing it at a single time: there are many
incompatible alternative frameworks or sample spaces. And the same principles
of Liberty, Equality, and Utility apply.  The physicist is at Liberty to
choose a sample space that is useful for discussing questions he considers
interesting; no one is more ``fundamental'' than another.  Two incompatible
spaces cannot be combined and, as we shall see in Sec.~\ref{sbct3.4}, there
are circumstances under which this single framework rule must be made more
stringent.  But not all sample spaces are equally useful.  As a trivial
example, the family $\FM_1$ introduced above is obviously more helpful than
$\FM_2$ if the physicist is trying to answer a question like: ``The silver
atom had $S_z = +1/2$ at time $t_3$; what can one say about $S_z$ at some
earlier time?''

\xb\outl{Rules for assigning probs are given implicitly in quantum textbooks.
}\xa

\xb\outl{Detailed discussion of how to do it consistently is in CQT
}\xa

Some of the rules needed for assigning probabilities to histories comprising a
particular family or framework in a closed quantum system are given in
textbooks in the particular case of the Born rule, Sec.~\ref{sbct3.3} below.
However, the presentation tends to be accompanied by references to
measurements and wave function collapse that do not properly reflect how
measurements are actually used and interpreted by competent experimental
physicists, the topic of Sec.~\ref{sct5} below.  As the appropriate way to
introduce probabilities in a \emph{closed} system is discussed in considerable
detail, with numerous examples, in CQT, the present discussion is limited to
the highlights.  We begin by considering the simplest situation, unitary
families, in Sec.~\ref{sbct3.2}; then two-time families, the traditional Born
rule, in Sec.~\ref{sbct3.3}.  These will serve to set forth the basic strategy
of the histories approach.  The extension of these ideas to more complicated
situations, which involves additional technical difficulties, is briefly
taken up in Sec.~\ref{sbct3.4}.

\xb
\subsection{Unitary families and the uniwave}
\label{sbct3.2}
\xa

\xb\outl{Unitary families. Probabilities 0 or 1. 
}\xa

The simplest type of history family for a closed system from the perspective
of assigning probabilities is one in which they are all either 1 or 0.  Such a
family is a \emph{unitary} family.  The simplest way to construct such a
family is to assume a particular pure state $\ket{\psi(t_0)}$ for the system
at an initial time $t_0$, and integrate Schr\"odinger's equation to obtain
$\ket{\psi(t_m)}$ at the later times of interest, which we assume constitute a
finite collection $t_0 < t_1 < \cdots t_f$.  At time $t_m$ introduce a PD with
just two projectors: $P_m = [\psi(t_m)]$ and its negation $I-P_m$, and
construct the history sample space by choosing one of these for each $t_m$.
This family satisfies the consistency conditions in Sec.~\ref{sbct3.4}, and if
$P_0$ at $t_0$ is assigned probability 1, the single history $P_0 \od P_1
\od\cdots P_f$ has probability 1 and all other members of the sample space
have probability 0.  Consequently, one can conclude that if the system has
property $P_0$ at $t_0$ then it has the property $P_m$ at each later time
$t_m$.  Alternatively, one can reach the corresponding result by conditioning
on $P_n$ at some particular time $t_n$.  Hence for a unitary family the
histories approach yields a deterministic quantum dynamics.

\xb\outl{Uniwave: unitary time evolution of universe
}\xa

\xb\outl{Uniwave in other interpretations of QM
}\xa

Let us introduce the technical term \emph{uniwave} for a a wave function
$\ket{\psi(t)}$ that satisfies Schr\"odinger's equation, thus undergoes a
\emph{unitary} time evolution, for a closed system, the \emph{``universe.''}
In many interpretations of quantum mechanics the uniwave, often referred to
simply as ``the wave function,'' plays a central role.  But not in the
histories approach; with the advent of Equality the uniwave is only a common
citizen and no longer king.  It describes reality for a unitary family,
but not in more general probabilistic situations where, as we shall see in
Sec.~\ref{sbct3.3}, it is further demoted to a mere mathematical tool or
pre-probability for calculating probabilities which can, if one wants, be
obtained using different solutions to Schr\"odinger's equation.
This is in contrast to some other interpretations of quantum mechanics.  The
uniwave is an absolute monarch in the Everett (many worlds) interpretation
\cite{Brrt09,Vdmn09}, where it constitutes the official ontology: at each time
it represents all of reality.  Bohmian mechanics \cite{Glds06} adds additional
(hidden) variables, but their motion in time is then determined by the the
uniwave.

\xb
\subsection{Born rule}
\label{sbct3.3}
\xa

\xb\outl{Families with two times.  Born rule stated
}\xa

After unitary families, in which the probabilities are trivial, the simplest
situation for a closed system is a family that involves only two times $t_0$
and $t_1$.  One typically assumes that $t_0$ is earlier than $t_1$, but this
is not essential.  Suppose the histories framework is provided by an
orthonormal basis $\ket{\psi_0^j}$ at $t_0$ and another, in general different,
orthonormal basis $\ket{\phi_1^k}$ at $t_1$.  Let $T(t_1,t_0)$ be the
corresponding unitary time development operator, given by $e^{-i(t_1-t_0)
  H/\hbar}$ in the case of a time-independent Hamiltonian H. The Born rule
assigns a weight
\begin{equation}
  |\mted{\phi_1^k}{T(t_1,t_0)}{\psi_0^j}|^2 = 
 |\mted{\psi_0^j}{T(t_0,t_1)}{\phi_1^k}|^2 
\label{eqn7}
\end{equation}
to the history $[\psi_0^j]\od [\phi_1^k]$, which can then be interpreted as the
conditional probability of $[\phi_1^k]$ at time $t_1$ given $[\psi_0^j]$ at time
$t_0$, or of $[\psi_0^j]$ at $t_0$ given $[\phi_1^k]$ at $t_1$.

\xb\outl{Born rule in textbooks: psi(t) as pre-probability
}\xa

The complete formal symmetry between the two times is not so obvious if one
writes the weight in \eqref{eqn7} in the equivalent form
\begin{equation}
  |\inpd{\phi^k}{\psi(t_1)}|^2,\quad 
 \ket{\psi(t)}:= T(t,t_0)\ket{\psi(t_0)},\quad
 \ket{\psi(t_0)}:= \ket{\psi_0^j}
\label{eqn8}
\end{equation}
for some particular choice of $j$.  That is, one starts with an initial state
at $t_0$, constructs the corresponding uniwave, and uses it to compute the
weight. In the histories approach there can be no objection to \eqref{eqn8} as
a mathematical formula, but since $\ket{\psi(t_1)}$, or more precisely its
projector $[\psi(t_1)]$, will in general be incompatible with the chosen PD
$\{[\phi_1^k]\}$ at time $t_1$, one cannot consistently speak of it as a
property of the system at $t_1$: it does not make sense within the family of
histories whose probabilities are being calculated by means of the Born
rule. Instead, $\ket{\psi(t_1)}$ is best thought of as a
\emph{pre-probability} in the notation of CQT: a mathematical tool used for
computing probabilities, but which need not have any counterpart in physical
reality.

\xb\outl{Uniwave not essential; weights by integrating backwards in time
}\xa

A further indication that in this situation the uniwave is playing a
subsidiary role and does not represent a physical property is that the
Born weights in \eqref{eqn7} can be calculated by an alternative route that
makes no reference to it. Thus let us define the kets
\begin{equation}
  \ket{\phi_0^k} = T(t_0,t_1) \ket{\phi_1^k},
\label{eqn9}
\end{equation}
i.e., for each $k$ integrate Schr\"odinger's equation ``backwards'' from $t_1$
to $t_0$, starting with $ \ket{\phi_1^k}$ as the ``initial'' state.  Then it
is obvious that
\begin{equation}
  |\mted{\phi_1^k}{T(t_1,t_0)}{\psi_0^j}|^2 = |\inpd{\psi_0^j}{\phi_0^k}|^2,
\label{eqn10}
\end{equation}
so we have obtained the weights in \eqref{eqn7} without the help of the
uniwave defined in \eqref{eqn8}, by employing the $\ket{\phi_0^k}$ as
pre-probabilities.

\xb\outl{Uniwave must be dethroned to solve measurement (and Schr cat) problem
}\xa

Dethroning the uniwave from king to the very subsidiary role of serving as a
pre-probability, with even less reality than a probability, solves the first
measurement problem. Instead of declaring that the pointer is in some
macroscopic superposition, use a framework in which the projectors in the PD
refer to different pointer positions, and the (first) measurement problem has
disappeared.  Equivalently, the Schr\"odinger cat paradox vanishes once the
physicist is given Liberty to choose an appropriate framework of the
quasiclassical form, Sec.~\ref{sct4}, for events at later times.

\xb\outl{Orthonormal bases not essential for Born rule; one can use coarser PD
}\xa

\xb\outl{Born rule makes no reference to measurements
}\xa

Two additional remarks.  First, it is not necessary to use orthonormal bases
at $t_0$ and $t_1$; coarser PD's are acceptable, e.g., $\{P^j\}$ at $t_0$ and
$\{Q^k\}$ at $t_1$, with the weight formula \eqref{eqn7} replaced with the
more general
\begin{equation}
  \Tr(Q^k T(t_1,t_0) P^j T(t_0,t_1)) =
 \Tr(P^j T(t_0,t_1) Q^k T(t_1,t_0)),
\label{eqn11}
\end{equation}
which reduces to \eqref{eqn7} when $P^j = [\psi^j]$ and $Q^k = [\phi^k]$.
The only thing one needs to be careful about is normalization when turning
weights into probabilities, thus
\begin{equation}
  \Pr(Q^k\vb P^j) = \Tr(Q^k T(t_1,t_0) P^j T(t_0,t_1))/\Tr(P^j).
\label{eqn12}
\end{equation}
Second, our discussion of the Born rule made no references whatever to
measurements: it applies to any family of histories in a closed system as long
as they involve just two times.  Naturally, the closed system may include a
measuring apparatus, and the Born rule applies to the whole system.  For more
on measurements, see Sec.~\ref{sct5}.

\xb
\subsection{Multiple times: consistency}
\label{sbct3.4}
\xa

\xb\outl{Born rule for two times; more general rule needed for more times
}\xa

The Born rule suffices for assigning probabilities or weights to histories
inside a close quantum system when \emph{only two times} are involved.  With
three or more times an additional \emph{consistency condition} is needed. To
see why, consider a family of histories involving the three times $t_0 < t_1 <
t_2$.  The appropriate version of \eqref{eqn7}, or more generally
\eqref{eqn11}, can be used to calculate weights for histories involving just
the two times $t_0$ and $t_1$, or just $t_0$ and $t_2$, or $t_1$ and $t_2$,
but there is no guarantee that these can be turned into a corresponding joint
probability distribution for events at all three times. This resembles a
situation in classical stochastic processes: a rule which relates
probabilities of events at only two times cannot be used to generate a
multitime probability distribution without making additional assumptions;
e.g., that the process is Markovian. The quantum situation is more subtle, and
resembles the problem faced when deciding what to do with incompatible PDs at
a single time, Sec.~\ref{sct2}.

\xb\outl{Decoherence functional, consistency condition, and weights
}\xa

\xb\outl{Born weights are obtained as a special case for two-time histories
}\xa

In the histories approach one first postulates a reasonable mathematical form
for weights appropriate to a family of histories of a closed quantum system
involving an arbitrary (finite) collection of times, and then only applies it
to families which satisfy a fairly stringent \emph{consistency condition}.
Both weights and consistency can be obtained from the \emph{decoherence
  functional}
\begin{equation}
 \DM(\al,\bt) = Tr[ K\ad(Y^\al)\, K(Y^\bt) ] 
\label{eqn13}
\end{equation}
defined in terms of a \emph{chain operator}
\begin{equation}
 K(Y^\al) := F_f^\al T(t_f,t_{f-1}) F_{f-1}^\al T(t_{f-1},t_{f-2}) \cdots
F_1^\al T(t_1,t_0) F_0^\al,
\label{eqn14}
\end{equation}
with the analogous definition for $K(Y^\bt)$.
The consistency condition is then the requirement
\begin{equation}
 D(\al,\bt) = 0\text{ for } \al\neq\bt,
\label{eqn15}
\end{equation}
and when it is satisfied probabilities can be computed from the collection of
nonnegative weights
\begin{equation}
 W(\al) = \DM(\al,\al).
\label{eqn16}
\end{equation}
When $f=1$, so the family of histories involves only the two times, the
consistency condition is always fulfilled, so it did not have to be considered
in our earlier discussion of the Born rule in Sec.~\ref{sbct3.3}, and the Born
weights coincide with those in \eqref{eqn16}.  In this sense the histories
approach is a generalization of the Born rule.

\xb\outl{Details in CQT; should not use weaker consistency condition given there
}\xa

For further details the reader is referred to Chs.~10 and 11 of CQT, and to
the chapters that follow for various applications that draw out the physical
significance of the consistency condition and probabilities which can be
assigned when it is satisfied.  That treatment needs to be updated in two
respects.  First, the consistency condition used throughout CQT is the one
stated above: the ``medium decoherence'' condition in the terminology of
\cite{GMHr93}.  A weaker condition mentioned in Sec.~10.2, but (fortunately!)
never employed there or elsewhere in CQT, has been shown to be unsatisfactory
by Diosi \cite{Dsi04}.  Second, anyone who wants to understand the
consistency conditions, or teach them to students, would be well advised to
start with the simpler situation represented by chain kets and discussed in
Sec.~11.6, rather than the very general form given in Ch.~10.

\xb\outl{Consistency conditions limit realm of discussion
}\xa

\xb\outl{They permit what physicists need; removes spurious nonlocality
}\xa

The use of consistency conditions extends the cautious and conservative
approach employed in Sec.~\ref{sbct2.1}: restrict one's discussion to cases
that make sense. In particular, use only those sample spaces of histories of a
closed quantum system for which the assignment of probabilities based upon the
dynamical laws can be done in a consistent fashion. If the family does not
satisfy the consistency conditions its discussion is deferred to the robots,
Sec.~\ref{sbct2.1}.  In the meantime the families which do satisfy the
consistency conditions seem adequate: what is allowed covers what
physicists need to talk about, and what is excluded includes things like the
mysterious instantaneous nonlocal influences that have long been an
embarrassment to quantum foundations and are often thought to be an impediment
to connecting quantum theory with special relativity; see Sec.~\ref{sct6}.
Though quite restrictive, consistency conditions permit a much wider
discussion than found in textbooks, or allowed by the black box strictures
of quantum orthodoxy.  But see the further comments in Sec.~\ref{sbct8.4.2}.

\xb\outl{"Framework" refers to CONSISTENT family. Single framework rule extended
}\xa

For the reasons just mentioned it has become customary when discussing the
stochastic time dependence of closed quantum systems to limit the concept of
``framework'' to families for which the consistency conditions are satisfied.
This means both a restriction on the form of the history space PD allowed for
consistent probabilistic reasoning about what is going on in a closed quantum
system, and an extension of the single framework rule to exclude combining two
consistent families when the resulting common refinement will not satisfy the
consistency conditions.  If they cannot be combined they are said to be
incompatible.

\xb\outl{Qm ontology: consistent families.
}\xa

\xb\outl{Mutual incompatibility understood the same way as for a system at one time
}\xa

Thus it seems reasonable to include in a consistent quantum ontology only
those families of histories for a closed system which satisfy consistency
conditions. The physicist will have many different incompatible families from
which to choose, and can approach this multiplicity with the same attitude
adopted in Sec.~\ref{sct2} for a system at a single time.  In a
given family one and only one history actually occurs: they are mutually
exclusive.  The choice of family is up to the physicist, and will generally be
made on the grounds of utility, e.g., how to model a particular experimental
situation with apparatus put together by a competent experimentalist.
This choice has no influence upon what really goes on in the world, and
alternative choices applied to the same set of data and conclusions will yield
consistent results.  The quantum world is of such a nature that it can be
described in distinct ways which (in general) cannot be combined into a single
all-encompassing description.

\xb
\subsection{Which history occurred?}
\label{sbct3.5}
\xa

\xb\outl{Which history actually occurred?
}\xa

Mixing up the concepts of quantum incompatibility and
mutual exclusivity has led to a serious misunderstanding of the ontology of
stochastic histories, which is a reflection of the problem of understanding
what is real about a quantum system at a single time.  Here the issue is:
which history actually occurred?

\xb\outl{In a single family, one and only one history occurred 
}\xa

The first step in the reply is to say that if a single framework, a single
consistent family, of histories is in view, the sample space, represented
mathematically by an appropriate PD of the history identity, is a collection
of mutually-exclusive possibilities, one and only one of which actually
occurs.  The same as in the case of three tosses of a coin: one and only one
of the eight possible sequences occurs on the particular occasion in which the
experiment is carried out.  (Naturally, members of the corresponding event
algebra need not be mutually exclusive: the event of tails on the first two
tosses is distinct from that of tails on the last two tosses, but the two are
not mutually exclusive.)
The same is also true for any ordinary (not quantum) scientific application of
probability theory: only one of the potential possibilities represented in the
sample space of histories actually occurs.
The notion that there are separate universes in which each possibility in the
sample space is realized seems rather bizarre when one is considering, e.g.,
whether it will snow in Pittsburgh on Christmas day in the year 2050.  In this
respect quantum ontology as understood using the histories approach is much
closer to other applications of probability theory than is the ``many worlds''
\cite{Brrt09,Vdmn09} understanding of Everett's ideas. 

\xb\outl{Two incompatible frameworks: cannot say pair of histories occurred
}\xa

The second step in the reply addresses the issue of two or possibly more
incompatible families of histories.  What has just been said about one and
only one history \emph{applies separately to each family}.  But
\emph{incompatible families} cannot be \emph{combined}.  Hence it is
meaningless (histories quantum mechanics assigns it no meaning) to assert that
there must be some pair of histories $(Y_1,Y_2)$, such that the first occurred
in family $\FM_1$ and the second in the incompatible family $\FM_2$.
Assertions of this sort are just what the single framework rule is designed to
exclude.  For an example of how this works in a particular gedanken experiment
see Sec.~18.4 of CQT.

\xb
\section{Classical Limit}
\label{sct4}
\xa

\xb
\subsection{Quasiclassical frameworks}
\label{sbct4.1}
\xa

\xb\outl{Introduction. Omnes and GMH.  No outstanding problems
}\xa

An important feature of a satisfactory quantum ontology is that it should
explain how classical mechanics at the level of macroscopic objects, such as
dust particles or pennies or planets, emerges from a fundamental quantum
description of the world.  This problem has been addressed from the histories
perspective, and at of the present time, while the task is still incomplete,
there are no outstanding problems that indicate difficulties in the way of an
approach pursued in somewhat different ways by Omn\`es \cite{Omns99,Omns99b}
and by Gell-Mann and Hartle \cite{GMHr93}. The following remarks provide only
a fairly elementary explanation of the basic approach.

\xb\outl{Enormous size of subspaces needed for quasiclassical description
}\xa

\xb\outl{Quasiclassical PD: not a precise term
}\xa

The first important idea is that of \emph{coarse graining} the Hilbert space
by using an appropriate PD in which the subspaces correspond to ordinary
macroscopic descriptions of the world.  Properties associated with a pointer
pointing in a particular direction, or its modern counterpart, a
(``classical'') bit represented by some macroscopic state of an electronic
memory device, are represented by subspaces of the Hilbert space of
exponentially large dimension $10^\nu$, where where $\nu$ is itself an
enormous number, e.g., $10^{10}$.  In such cases many different coarse-grained
projectors could be used to represent what a physicist would be say is
essentially the same macroscopic property.  What seems plausible is that a
suitably chosen coarse-grained \emph{quasiclassical} PD can represent the
different possibilities in a macroscopic situation, such as the outcome of a
laboratory measurement, with sufficient accuracy ``for all practical
purposes''.
A quasiclassical PD both cannot and should not be precisely defined; there
will be many different PDs that provide very similar results.  If this seems
sloppy by the standards of pure mathematics or philosophy, the appropriate
response is that approximations are in practice essential in all of physics,
and which approximations are appropriate in any particular case are to some
extent a matter of judgment, and depend upon the problem being addressed.

\xb\outl{Quasiclassical framework of histories
}\xa

\xb\outl{Classical dynamics emerging from stochastic Qm dynamics
}\xa

The next step is to choose a family of quasiclassical histories, a
\emph{quasiclassical framework}, in which events at each time correspond to a
quasiclassical PD.  In constructing this framework the time intervals should
not be made too small, as a judicious choice will make it easier to satisfy
the consistency conditions, Sec.~\ref{sbct3.4}, for the probabilistic dynamics
of a closed system.  What one expects is that given a suitable coarse graining
of both phase space and time, classical dynamics will emerge as an
approximation to the more exact, but extremely difficult to calculate, quantum
dynamics. This has not been proved, but various calculations and arguments
make the existence of such a framework (or frameworks, since the choice is not
unique) plausible.  In this connection several comments are in order.

\xb\outl{Qm dynamics can yield results which are close to deterministic
}\xa

First, quantum dynamics, as already noted in Sec.~\ref{sct3}, is not to be
thought of as unitary time development, even though the latter is always a
possible description for a closed system.  Instead it is fundamentally
stochastic.  How can a deterministic classical mechanics emerge from an
underlying stochastic quantum dynamics?  This depends on the magnitudes of the
probabilities.  It is easy to envisage situations in which given a suitable
coarse graining the probability of one macro property being followed by
another a short or even much longer time later is almost equal to 1, with the
deviation being very small. This then amounts in practice to a deterministic
dynamics.

\xb\outl{Continuous Cl time vs discrete Qm time
}\xa

Second, classical mechanics is formulated in terms of a continuous, not a
discrete time development.  How can this be reconciled with a quantum
stochastic dynamics in which time intervals need to remain finite? Again, the
time intervals of interest when considering the flight of a golf ball are
quite large compared with the time scales which enter into the quantum physics
of matter.  What one expects is that a discrete time development consistent
with quantum consistency conditions will in many circumstances lead to a time
dependence which can be satisfactorily \emph{approximated} by the differential
equations of classical mechanics. This seems a reasonable hope given what we
understand at present about the quantum mechanics of large systems, even if 
much is still waiting to be confirmed by serious and systematic studies.

\xb\outl{Radioactive decay
}\xa

\xb\outl{Cl mechanics in situations of chaos is close to indeterministic
}\xa

Third, in addition to situations in which a continuous and deterministic
classical dynamics should be a good approximation to the quantum world, there
will also be situations in which the intrinsic uncertainties associated with
quantum dynamics manifest themselves as genuinely stochastic behavior at the
macroscopic level, behavior that is quite unpredictable in terms of some
earlier state of the universe.  Radioactive decay, detected by devices whose
later behavior can be fitted within a quasiclassical framework, is an obvious
example.  Another is chaotic motion of systems in which, according to
classical mechanics itself if taken literally, any small change in the
starting point in the system's phase space is rapidly magnified until for all
practical purposes the motion is no longer deterministically linked to an
initial state.  One would hardly expect a quasiclassical quantum description
applied to such a situation to yield a deterministic outcome, and the
stochastic quantum dynamics of the histories approach provides a plausible
structure in which such a probabilistic time development can be seen to be
consistent with a fully quantum mechanical world.  Calculations to verify
these ideas face formidable technical problems, but nothing in our present
state of knowledge suggests there is any difficulty in principle in describing
such situations in quantum terms.

\xb\outl{Issue of consistency conditions being fulfilled. Dowker and Kent
}\xa

Fourth, it is reasonable to expect that the consistency conditions,
Sec.~\ref{sbct3.4}, will be satisfied in a quasiclassical framework. Although
when expressed mathematically the conditions \eqref{eqn15} are quite strict,
the attitude of the physicist is that as long as the violations of these
conditions are small compared to the diagonal weights \eqref{eqn16} one is
interested in, it does not matter.  That this attitude is not unreasonable is
supported by the work of Dowker and Kent \cite{DwKn96}, who on the basis of a
parameter counting argument concluded that in the vicinity of an
almost-exactly consistent family there is another one, obtained by slightly
adjusting the properties under discussion, that is to say the projectors
representing them, which is fully consistent, with the mathematical conditions
exactly satisfied.  Since in the case of quasiclassical coarse grainings there
is enormous room (as viewed microscopically) to make adjustments which leave
macroscopic properties the same for all practical purposes, consistency
conditions do not seem to pose a significant problem.

\xb\outl{Decoherence can render certain histories consistent
}\xa

Fifth, what is the role of decoherence?  Up till now most research on this
topic, see for example \cite{Jsao03,Zrk03} has been carried out without an
adequate ontological framework. Hence there are doubts as to the nature of the
appropriate concepts, and what the calculations actually mean.  Simple models
suggest that decoherence in the form of a fairly weak interaction with a
suitable environment can render certain histories of a system consistent when
they would otherwise, in the absence of the environment, be inconsistent.
This supports the plausibility of classical physics emerging from quantum
physics in an appropriate quasiclassical framework.  In the histories approach
decoherence is \emph{not} needed to resolve the measurement problem; see the
following section.

\xb
\subsection{Persistence of quasiclassical behavior}
\label{sbct4.2}
\xa

\xb\outl{ Dowker and Kent: quasiclassicality need not persist
}\xa

Since it has been frequently cited, e.g. \cite{Schl04,Wllc08}, in work which
makes no reference to the detailed rebuttal found in \cite{Grff98}, it seems
worthwhile discussing a criticism of the histories approach found in
\cite{DwKn96}, and also in \cite{Knt96,Knt98}, having to do with persistence
in time of quasiclassical behavior: Suppose we know (or suppose) that the
world has been quasiclassical up to now.  How do we know that
quasiclassicality will continue to be the case tomorrow?  There is no
guarantee (so the criticism goes) that this will be the case, because the
histories approach contains no principle which singles out the correct family
for describing the world, and it is easy to construct examples of consistent
families which employ quasiclassical PDs up to some point in time, but then at
later times use PD's which are incompatible with the quasiclassical form.

\xb\outl{quasiclassical is a type of description, not a property of the world
}\xa

\xb\outl{analogy of using coarse graining to derive hydrodynamic laws
}\xa

The basic misunderstanding arises from supposing that ``quasiclassical'' is an
adjective that somehow describes, within histories quantum mechanics, some
property of the world.  But it is not a property of the world; instead it is a
property of a certain type of \emph{description} of the world.  The analogy of
coarse graining a classical phase space in order to derive, say, hydrodynamics
from classical statistical mechanics can help show where the error lies.  This
coarse graining has worked successfully for deriving the hydrodynamic laws
that apply up to time $t_1$.  But what guarantees that the same coarse
graining will apply at times later than $t_1$?  Only the fact that the
physicist who has carefully done the work is not likely to throw away an idea
that has succeeded and replace it with something that doesn't work as well.
This, of course, has nothing to do with the flow of water through real pipes;
the laws of hydrodynamics do not change because the physicist trying to relate
them to microscopic (classical) mechanics has adopted an ineffective approach.

\xb\outl{Frameworks are not mutually exclusive; no law of nature gives the right one
}\xa

To put it in a different way, the relationship between different frameworks is
not one of mutual exclusivity, where one is right and the other is wrong. Once
one appreciates this fact, there is no reason to look for a law of nature that
singles out a particular framework.  The existence of alternative
frameworks in no ways invalidates the conclusions drawn using a particular
framework.  Had Dowker and Kent been able to conclude that \emph{no}
quasiclassical framework could be consistently used over the long period of
time required to get from, let us say, the big bang to a few billion years
into the future, that would have counted as a serious objection to the
histories program.  But that is not not the case.

\xb
\section{Preparation and Measurement}
\label{sct5}
\xa

\xb\outl{Preparation and measurement as quantum processes
}\xa

The black box approach to quantum theory mentioned in Sec.~\ref{sct1} makes
reference to ``preparation'' and ``measurement'' as unanalyzable (from a
quantum perspective) macroscopic processes to be understood in classical terms,
while talk about what it is that is actually prepared, or what a measurement
actually measured, is forbidden, at least by the strict rules of this
particular orthodoxy.  The consistent quantum ontology that is the subject of
this paper makes it possible to open the black box without falling into the
quantum foundations swamp, and one can ``watch'' what is going on
inside. But in addition it provides the tools needed to treat both preparation
and measurement as examples of \emph{quantum mechanical} processes governed by
exactly the same principles that apply to all other quantum processes,
microscopic or macroscopic, whether or not they have anything to do with 
either  preparation or measurement.

\xb\outl{References to CQT chapters on measurements
}\xa

Measurements have been extensively discussed in CQT Chs.~17 and 18, to which
we refer the interested reader for details on topics such as macroscopic
description of measurements in which one does not assume the apparatus is in a
pure state prior to measurements, and the role of thermodynamic
irreversibility.  These are omitted from the following discussion in order to
focus on the essentials. We also omit discussions of delayed choice
experiments, Ch.~20 of CQT, which have sometimes been misinterpreted to mean
that in the histories approach the future can influence the past.

\xb\outl{What follows: simple model discussions of prep, measurement
}\xa

\xb\outl{plus a description of what goes on inside the black box
}\xa

In what follows we shall use a simple model, a slight modification
of the one introduced by von Neumann, in order to discuss
both preparations and measurements, starting with the latter.  Then we shall
argue that when the appropriate analyses are put together one arrives at
a realistic quantum description of at least some aspects of what goes on 
at a microscopic level between a preparation and a measurement, thus opening
the black box.

\xb
\subsection{Measurement model}
\label{sbct5.1}
\xa

\xb\outl{Modified vN model (final ket for system is always the same)
}\xa

\xb\outl{times $t_0 < t_1 < t_2$; measurement interaction between $t_1$ and $t_2$
}\xa

Here is the modification of von Neumann's model. Let $\HM_s$ be the Hilbert
spaces associated with the system to be measured---hereafter referred to as
the \emph{particle}---and $\HM_M$ that of the measuring apparatus,
including its environment if that is of interest.  The two taken together form
a closed quantum system, so unitary time development of the combination
makes sense.  We assume that at time $t_0$ the particle is in a normalized
state
\begin{equation}
  \ket{\psi_0} = \sum_j c_j\ket{s^j},
\label{eqn17}
\end{equation}
a linear combination of orthonormal basis states $\{\ket{s^j}\}$ of $\HM_s$,
and the apparatus is in the state $\ket{M_0}$.  Let $t_0 < t_1$ be two times
preceding the time interval from $t_1$ to $t_2$ during which the particle
interacts with the apparatus, and let the unitary time development from $t_0$
to $t_1$ to $t_2$ be given by
\begin{equation}
  \ket{s^j} \ot \ket{M_0} \ra \ket{s^j} \ot \ket{M_1}  \ra
\ket{s^1} \ot \ket{M^j}.
\label{eqn18}
\end{equation}
Note that in \eqref{eqn18} at the final time $t_2$ the particle is always in
the \emph{same} state $\ket{s^1}$ whatever may have been the earlier state $
\ket{s^j}$.  In words, during the time interval from $t_0$ to $t_1$ the
particle state does not change, whereas the apparatus undergoes a time
evolution $\ket{M_0} \ra \ket{M_1}$, the nature of which is unimportant for
the present discussion as long as $\ket{M_1}$ is an appropriate ``ready''
state that leads at time $t_2$ to a state $\ket{M^j}$ corresponding to the
pointer being in a definite direction that is determined by the earlier state
$\ket{s^j}$ of the particle at time $t_1$.  One could, for example, think of a
particle with a spin degree of freedom moving towards an apparatus where some
component of the spin will be measured. In this case the $\{\ket{s^j}\}$ would
refer to the spin states, whereas the center of mass motion of the particle
should be thought of as part of the apparatus, and the change from $\ket{M_0}$
to $\ket{M_1}$ could incorporate the center of mass motion. As noted in
Sec.~\ref{sct4}, distinct macroscopic states always correspond to orthogonal
subspaces (in an appropriate quasiclassical framework) of the apparatus
Hilbert space, so there is no problem in supposing that the $\{\ket{M^j}\}$ in
\eqref{eqn18} are orthogonal, as required by unitary time evolution.

\xb\outl{Unitary evolution when system is in initial superposition state
}\xa

By linearity the dynamics in \eqref{eqn18} applied to the initial state
in \eqref{eqn17} results in a unitary time evolution
\begin{equation}
  \ket{\Psi_0} = \ket{\psi_0} \ot \ket{M_0} \ra 
\ket{s^1} \ot (\sum_j c_j\ket{M^j}),
\label{eqn19}
\end{equation}
from $t_0$ to $t_2$, where Schr\"odinger's fearsome cat, equivalently the
first measurement problem, appears on the right side.  Its resolution is
straightforward: in order to have something useful for comparison with
the work of a competent experimentalist, we should employ a framework in which
the different pointer positions make sense, a quasiclassical framework of
the type mentioned in Sec.~\ref{sct4}.  Let its projectors $\{P^j\}$
form a PD on $\HM_M$ chosen in such a way that
\begin{equation}
  P^j\ket{M^j} =\ket{M^j}.
\label{eqn20}
\end{equation}
(As usual the symbol $P^j$ can also stand for $I_s\ot P^j$.)

\xb\outl{Born rule for probabilities of pointer positions
}\xa

Then a straightforward application of the Born rule, \eqref{eqn11} and
\eqref{eqn12} in Sec.~\ref{sbct3.3}, to this situation results in
\begin{equation}
  \Pr(P^j \text{ at }t_2) = |c_j|^2,
\label{eqn21}
\end{equation}
the standard result found in every textbook.  This is actually a probability
conditional upon the initial state $\ket{\Psi_0}$ at $t=0$, but in this and
the what follows we simplify the notation by omitting this condition, which is
always the same.  Note that there has been no reference to a distinct type of
time evolution, as one finds in von Neumann's original treatment.  Instead we
have simply applied the principles of stochastic quantum time evolution
discussed in Sec.~\ref{sct3} to a particular situation involving a closed
quantum system that includes the measuring apparatus along with the measured
particle.  The first measurement problem has vanished, or one might say that
it never makes its appearance once Equality has terminated the absolute reign
of the uniwave.  There is, to be sure, an alternative unitary framework in
which the fearsome cat, the right side of \eqref{eqn19}, is present at $t_2$,
and physicists, philosophers and science fiction writers are at Liberty to
contemplate it as long as they keep in mind its incompatibility with (and thus
irrelevance to) the sorts of descriptions commonly employed by competent
experimental physicists when describing work carried out in their
laboratories.

\xb\outl{Correlations handle the second measurement problem
}\xa

Now that we have a description in which the pointer has stopped wiggling, so
its position makes sense, we can address a second concern of the competent
experimentalist: how are these pointer positions related to the state of the
particle just before the measurement took place?  He has, after all, designed
the apparatus so that a pointer position $P^j$ corresponds to a prior state
$\ket{s^j}$ of the particle; is this consistent with a proper quantum
description of what is going on?  To address this question we need the more
detailed description provided by the family
\begin{equation}
  [\Psi_0] \odot \{[s^j]\} \odot \{P^k\},
\label{eqn22}
\end{equation}
where the alternatives at times $t_1$ and $t_2$ are enclosed in braces
$\{\}$. These histories involve three times, so consistency is important, but
that is easily checked, see CQT Ch.~17 for details, and one arrives at 
\begin{equation}
  \Pr(s^j \text{ at } t_1 \text{ \AND\ } P^k \text{ at } t_2) =
\dl_{jk} |c_j|^2,
\label{eqn23}
\end{equation}
whence it follows, assuming that $|c_k|^2$ is positive so that the
conditional probability makes sense,
\begin{equation}
  \Pr(s^j \text{ at } t_1\vb 
  P^k \text{ at }  t_2) = \dl_{jk}.
\label{eqn24}
\end{equation}
In words, given the pointer position is $P^k$ at time $t_2$, it follows with
certainty (conditional probability 1) that the particle was in the
corresponding state at $t_1$, i.e., the apparatus was functioning according to
design.  Thus the second measurement problem has been solved.  This is, to be
sure, a very simple measurement setup, but it indicates an approach which can
be extended to more complicated situations.

\xb\outl{Comments on solution to second measurement problem
}\xa

\xb\outl{1. Role of little uniwave
}\xa

Several comments can be made about this result.  First, the solution to the
second measurement problem resembles that of the first in that a key role is
played by dethroning the uniwave.  In this case the uniwave of interest is not
the big one describing the very large universe of the particle-plus-apparatus,
but the small system of the particle in isolation as it travels through
orthodoxy's black box.  Strict orthodoxy, let it be noted, does not ascribe
reality to this little uniwave; it is simply a calculational tool.  The
histories approach, by contrast, is willing to treat it as real provided a
framework has been adopted in which it fits.  However, since at the time $t_1$
just before the measurement $\ket{\psi_0}$ is incompatible with the PD
corresponding to the properties $\{\ket{s^j}\}$ the apparatus has been
designed to measure, a framework in which $\ket{\psi_0}$ makes sense will not
be useful for discussing the measurement \emph{as a measurement}, i.e., as
measuring \emph{something}, so the physicist interested in that aspect of
things must use something else.

\xb\outl{2. Measurement outcome not correlated to later particle property, contra vN
}\xa

Second, it is worth noting that the measurement \emph{outcome} in this model,
the final $\ket{M^j}$ or $P^j$, lacks any interesting connection with the
state of the particle \emph{after} the measurement is over.  We used
$\ket{s^1}$ in the final term in \eqref{eqn18}, but any other choice would
have been equally good, including a final state $\ket{s^k}$ with the
relationship of $k$ to the earlier $j$ chosen randomly, or even von Neumann's
original version with $\ket{s^j}$ left unchanged in the time step from $t_1$
to $t_2$.  In the majority of situations in the laboratory in which
microscopic properties are measured by some macroscopic apparatus, the
particle after the measurement is in a state that is very different from the
one preceding the measurement.  As the measurement was designed to determine
the state of the particle before it was measured, its final state is entirely
irrelevant to the interpretation of the experiment.  There is no reason to
believe that the spin of a silver atom flying through a Stern-Gerlach remains
the same when it becomes attached to a glass plate; indeed, the very concept
of ``the spin of this silver atom'' becomes rather doubtful in that situation.
Von Neumann should not be faulted; his proposal came in the early days of
quantum mechanics when the role of measurement was not properly understood,
and it was valuable to have a model in which a second measurement could
confirm the outcome of the first.  But it is most unfortunate that quantum
textbooks continue to confuse students by treating the von Neumann model
together with its mysterious wave function collapse as somehow essential for
understanding the measurement process, a full quarter century after the
publication of a much better approach.

\xb\outl{3. Decoherence not needed
}\xa

Third, nowhere in the above discussion has any reference been made to
\emph{decoherence}.  This is because the appeal to decoherence is neither
necessary nor sufficient for resolving the measurement problem(s), despite
occasional claims in the literature that it can do so, see
e.g. \cite{Zrk03,Zrk09}, or at least that it plays some essential role. 
As has occasionally been pointed out by critics, e.g., \cite{Adlr03,Ghrr09},
decoherence does nothing to remove the underlying difficulty caused by the
assuming that quantum reality is represented by the uniwave. By contrast,
the histories approach, rather than appealing to decoherence, goes right to
the heart of the difficulty and resolves it by dethroning the uniwave.  This
is not to say that decoherence is unimportant for understanding the real
world, including real laboratory experiments. But its importance is not in
solving the measurement problems---and in any case it can make no pretense of
solving the second measurement problem---but rather for formulating and
justifying the quasiclassical descriptions and approximations needed to derive
ordinary macroscopic properties and classical dynamics from quantum theory, as
discussed in Sec.~\ref{sct4}.

\xb\outl{Frameworks needed for description of measurement as measurement
}\xa

Finally it is worth emphasizing that the correct description of a quantum
measurement \emph{as a measurement} requires that the framework contain both
an appropriate PD for the pointer positions at a time after the measurement
takes place \emph{and} a (different) PD representing the appropriate particle
properties \emph{before} it takes place.  Both requirements will in general be
incompatible, though in somewhat different ways, with the claim that the
uniwave constitutes quantum reality.  This helps explain why approaches based
on the uniwave have not succeeded in resolving the infamous measurement
problem of quantum foundations, and suggests that they are unlikely to do
any better in the future.

\xb
\subsection{Preparation model}
\label{sbct5.2}
\xa

\xb\outl{vN's model best viewed as preparation
}\xa

\xb\outl{Model analyzed using appropriate framework
}\xa

Von Neumann's original model, in which the final step in \eqref{eqn18} is
\begin{equation}
  \ket{s^j} \ot \ket{M_1}  \ra \ket{s^j} \ot \ket{M^j},
\label{eqn25}
\end{equation}
i.e., the particle state $\ket{s^j}$ is left unchanged by its interaction with
the apparatus, represents a nondestructive measurement, and can also be viewed
as a model for \emph{preparation}.  We again assume a starting state
$\ket{\psi_0}$ of the particle to be a superposition of the basis states,
\eqref{eqn17}, so unitary time development now leads, in place of
\eqref{eqn19}, to
\begin{equation}
  \ket{\Psi_0} = \ket{\psi_0} \ot \ket{M_0} \ra \sum_j c_j 
\Bigl(\ket{s^j} \ot\ket{M^j}\Bigr).
\label{eqn26}
\end{equation}
Again, unitary time evolution leads to a macroscopic superposition, and again
the route to a simple physical interpretation is to employ a PD $\{P^j\}$ of
the same type used previously for the apparatus, with \eqref{eqn20} satisfied,
i.e., each $P^j$ identifies a pointer state.  In addition we let $[s^i]$ be
the projector onto state $\ket{s^i}$ of the particle, so the total PD on the
tensor product of $\HM_s\ot\HM_M$ at the final time $t_2$ is $\{[s^i]\ot
P^j\}$. A simple application of the Born rule then leads to
\begin{equation}
  \Pr( [s^i] \text{ \AND\ } P^j \text{ at } t_2)
= \dl_{ij} |c_j|^2,
\label{eqn27}
\end{equation}
where, as previously, we omit explicit mention of the condition $\ket{\Psi_0}$.
It follows immediately, assuming $|c_j|^2$ is positive,
that 
\begin{equation}
  \Pr( [s^i] \text{ at } t_2 \vb P^j \text{ at } t_2) = \dl_{ij}.
\label{eqn28}
\end{equation}

\xb\outl{Comparison with wave function collapse
}\xa

In words, if after its interaction with the particle the apparatus is in the
state $P^j$, i.e., the pointer is in the direction $j$, then the particle is
in the corresponding state $[s^j]$, equivalently $\ket{s^j}$ (the phase does
not matter). This is consistent with von Neumann's idea that an outcome $P^j$
``collapses'' the wave function to $\ket{s^j}$, but rather than having to
invoke some ``magical'' collapse process the result in \eqref{eqn28} is a
quite straightforward consequence of using standard probabilistic reasoning.
Wave function collapse as used in textbooks is thus nothing but a system for
calculating conditional probabilities, and the books would be much less
confusing and less liable to misinterpretation were it stated as such.  Or,
given the enormous confusion wave function collapse has caused in quantum
foundations studies, it might be better to never mention the idea, and simply
teach students the proper use of probability theory, a subject which ought to
be included in every introductory quantum textbook.

\xb\outl{More general preparation: system states need not be orthogonal
}\xa

While this discussion has been limited to a very simple model of a
preparation process, it indicates the correct direction to go in more
complicated situations. Suppose, for example, that in place of \eqref{eqn26} 
there is a unitary time development
\begin{equation}
 \ket{\Psi_0} \ra \sum_j c_j 
\Bigl(\ket{r_j} \ot\ket{M^j}\Bigr),
\label{eqn29}
\end{equation}
starting with an initial state $\ket{\Psi_0}$ whose details need not concern
us, where the normalized states $\{\ket{r_j}\}$ need not be orthogonal.
However, the product states $\ket{r_j} \ot\ket{M^j}$ are orthogonal for
different $j$; so one has an example of dependent or contextual events or
properties as discussed in Ch.~14 of CQT. Once again, given the macroscopic
outcome $P^j$ one can conclude that the system is in the state $\ket{r_j}$,
regarded as a contextual property which depends on $P^j$, and there is no need
for wave function collapse, though that can be a convenient calculational
tool.

\xb\outl{Preparation followed by later measurement
}\xa

Finally, note that a description of a process of preparation followed by a
measurement, of course using separate pieces of apparatus, can be constructed
by combining the results presented here with those in Sec.~\ref{sbct5.1}.
Doing so effectively pries open orthodoxy's black box and allows the physicist
to ``see'' what has been prepared, and also understand the microscopic state
of affairs that gives rise to the final measurement outcome.  Events at
intermediate times while the particle is isolated can also be discussed, but
now one needs to pay attention to consistency requirements in order to arrive
at a sensible probabilistic description.

\xb
\subsection{POVMs}
\label{sbct5.3}
\xa

\xb\outl{Definition of POVMs; contrast with PDs
}\xa
	
In Sec.~\ref{sct2} it was noted that quantum properties of a system at a
single time are associated with a projective decomposition (PD): a set of
mutually orthogonal projectors on the Hilbert space that sum to the identity.
There is a more general notion of a decomposition of the identity, usually
referred to as a POVM (positive operator valued measure): a collection
$\{R_k\}$ of positive operators (Hermitian with nonnegative eigenvalues) which
sum to the identity.  Since the square of a positive operator is in general
not equal to itself we shall use the subscript $k$ as a label, while retaining
the superscript label for a PD. We assume that both the number of operators in
the POVM and dimension of the Hilbert space on which they act are finite.

\xb\outl{Physical interpretation of POVM
}\xa

A POVM is a useful mathematical tool in various situations, so it is natural
to ask whether it has some physical interpretation.  In the case of a PD,
which is a special case of a POVM, the elements are associated with mutually
exclusive properties, but for a general POVM such an interpretation is not
possible, especially when, as is usually the case, the different operators do
not commute with each other.  When used in some sort of ``measurement
situation'' the POVM typically serves as a pre-probability: a device for
calculating probabilities, but which does not itself have any direct physical
interpretation.  So when used this way a POVM does not have an ontological
reference. A simple example will serve to illustrate this point.

\xb\outl{Specific example of use of POVMs: measurement using ancillary system
}\xa

Consider a Hilbert space $\HM = \HM_s\ot\HM_A$, where $\HM_s$ is a system in
an unknown state $\ket{\psi_0}$ at $t_0$, whereas the auxiliary or
``ancillary'' system $\HM_A$ is known to be in a specific state $\ket{A_0}$;
thus the initial state of $\HM$ is
\begin{equation}
  \ket{\Psi_0} = \ket{\psi_0}\ot \ket{A_0}.
\label{eqn30}
\end{equation}
Next let $\{P^j\}$ be a PD for $\HM$ at a time $t_1 > t_0$, and assume the
time development from $t_0$ to $t_1$ is trivial, $T(t_1,t_0) = I$.  (One could
replace this with an arbitrary unitary without changing the following
discussion in any essential way.)  The probability of $P^j$ at time $t_1$ is
then given by the Born rule: 
\begin{equation}
  \Pr(P^j \text{ at } t_1 ) =
 \Tr(P^j [\Psi_0]) = \Tr_s (R_j [\psi_0])
\label{eqn31}
\end{equation}
where
\begin{equation}
  R_j := \Tr_A( P^j [A_0])
\label{eqn32}
\end{equation}
is an operator on $\HM_s$, that is, on the Hilbert space that holds the
unknown $\ket{\psi_0}$, and $\Tr_s$ and $\Tr_A$ denote partial traces.  It is
easily checked that the operator $R_j$ is positive and that $\sum_j R_j =
I_s$, so $\{R_j\}$ is a POVM on $\HM_s$.

\xb\outl{POVM provides convenient way to find Pr( physical property )
}\xa

What \eqref{eqn31} tells us is that the probability of a physical property
$P^j$ of the joint system represented by tensor product $\HM_s\ot\HM_A$ can be
calculated by means of a formula, the right side of \eqref{eqn32}, which only
involves the initial state $\ket{\psi_0}$ on the smaller Hilbert space.  To be
sure the initial state $\ket{A_0}$ of the ancillary system is playing a role,
as is evident from \eqref{eqn32}.  Nevertheless, in some circumstances the
POVM $\{R_j\}$ provides an efficient way of calculating probabilities, in a
way roughly analogous to the uniwave $\ket{\psi(t)}$ when used in the Born
formula in \eqref{eqn8}, or---perhaps a closer analogy---the
$\{\ket{\phi_0^k}\}$ in \eqref{eqn10}.  But in none of these cases is one
obliged to identify the pre-probability, which is a calculational tool, with
some aspect of physical reality; the probabilities of interest can very well
be computed by alternative methods in which the pre-probability never appears.

\xb\outl{In some other situations POVM might stand for a physical property
}\xa

To be sure, sometimes the same mathematical object which within one framework
serves only as a (dispensable) pre-probability can in another framework
represent a physical property; in particular, the uniwave can be used in this
way in unitary families.  This might in some circumstances be possible for
POVM elements, but it is not obvious, at least in general, how to construct
the appropriate framework.

\xb
\section{Quantum Locality }
\label{sct6}
\xa

\xb\outl{Claims that QM is nonlocal because of Bell inequality violations
}\xa

One frequently encounters the claim that quantum theory is intrinsically
nonlocal because of the presence of mysterious long-range influences which can
act instantaneously, in contraction to special relativity, but which cannot
carry any information, which means they are experimentally undetectable.%
\footnote{For a lengthy list of work by some of the
  principal advocates and critics, see the bibliography in \cite{Grff11}.} 
And sometimes it is asserted in addition that any future theory which yields
certain quantum mechanical results that that have been confirmed by
experiments must be similarly nonlocal (e.g., \cite{Mdln10}).  Such
claims are often based on observed violations of Bell's inequality, though one
can arrive at similar conclusions using Hardy's or the GHZ paradox.

\xb
\subsection{Genuine nonlocality}
\label{sbct6.1}
\xa

\xb\outl{Wave packet illustrates genuine Qm nonlocality
}\xa

The first point to be made is that there is a very genuine sense in which
quantum mechanics is nonlocal, already evident in a wavepacket $\psi(x)$ for a
particle in one dimension, which for convenience we assume is a continuous
function. If one regards
the corresponding $\ket{\psi}$ or projector $[\psi]$ as representing a quantum
property, it is easy to show that it fails to commute with the projector
$X(x_1,x_2)$ defined by
\begin{equation}
  (X(x_1,x_2) \psi) (x) = 
\begin{cases}
 \psi(x) & \text{if $x_1 \leq x \leq x_2$,}\\
 0      & \text{otherwise},
\end{cases}
\label{eqn33}
\end{equation}
unless the support of $\psi(x)$, the set of points where it is nonzero,
falls entirely inside or entirely outside the interval from $x_1$ to $x_2$.
Since the physical interpretation of the
property represented by $X(x_1,x_2)$ is that the particle lies inside this
interval, we have here a simple example of a sense in which a quantum particle
under certain circumstances can be said to be ``nonlocal,'' meaning that it
lacks a precise location.

\xb\outl{Assertion that Qm particle can be in two places at once is erroneous
}\xa

It is perhaps worth pointing out in this connection an error one frequently
encounters in popular expositions of quantum mechanics, though sometimes also
in more technical publications, where it is asserted that a quantum particle
``can be in two places at the same time.''  This is quite wrong, or at the
least thoroughly misleading, since the product of two projectors corresponding
to the particle lying in two nonoverlapping regions will be zero,
corresponding to the fact that this property is always false.  It would be
much better to say that the particle does not have a definite location.

\xb\outl{Spin singlet state: incompatible with all local properties
}\xa

A second and technically simpler example of such nonlocality, since it
involves only a finite-dimensional Hilbert space, is provided by the
well-known singlet state
\begin{equation}
  \ket{\psi_0} = \bigl(\ket{z_a^+,z_b^-} -\ket{z_a^-,z_b^+}\bigr)/\st 
\label{eqn34}
\end{equation}
of two spin-half particles $a$ and $b$.  It is a straightforward exercise to
show that the corresponding projector $[\psi_0]$ does not commute with any
nontrivial projector referring to particle $a$ alone, i.e., of the form $P\ot
I_b$ (the trivial projectors are 0 and the identity), or with one referring
to particle $b$ alone.  If particles $a$ and $b$ are in different locations
then one can say that $[\psi_0]$ is incompatible with any local property.  
While this is true, it is well to remember that what is involved is the
fundamentally quantum idea of incompatibility, which can have local as well as
nonlocal manifestations.  Thus $\ket{\psi_0}$ is the spin part of the ground
state wave function of the hydrogen atom, so one cannot consistently combine
the assertion that the hydrogen atom is in its ground state with (nontrivial)
talk about the spin angular momentum of either the electron or the proton.  In
this case the electron and the proton are not at separate locations. (Claiming
that they are would be inconsistent with the spatial part of the ground state
wave function.)

\xb
\subsection{Spurious nonlocality}
\label{sbct6.2}
\xa

\xb\outl{Spurious nonlocality discussed elsewhere
}\xa

\xb\outl{Charlie sends colored slips to Alice and Bob
}\xa

Having discussed cases of genuine quantum nonlocality, let us now turn to the
source of the mistaken notion that the quantum world is somehow pervaded by
nonlocal influences, which even their proponents admit cannot be used to
transmit information, and are hence experimentally undetectable.  As the whole
topic has recently been discussed at some length elsewhere \cite{Grff11}, it
will suffice for present purposes to focus on the central point where claims
of such nonlocality go astray.
Let us start by briefly repeating an example from \cite{Grff11}. Charlie
in Chicago takes two slips of paper, one red and one green, places them in two
opaque envelopes, and after shuffling them so that he himself does not know
which is which, addresses one to Alice and the other to Bob, who live in
different cities, but know the protocol Charlie is following.  Upon
receipt of the envelope addressed to her Alice opens it and sees a red slip of
paper.  From this she can immediately conclude that the slip in Bob's envelope
is green.  Her conclusion is not based on a belief that opening her envelope
to ``measure'' the color of the paper inside has some magical long-range
influence on what is in Bob's envelope.  Instead Alice employs statistical
reasoning in the following way.
From her knowledge of the protocol Alice can assign a probability of 1/2 to
each of the two possibilities: $A\colo G$ \AND\ $B\colo R$, green slip sent
to Alice and red slip to Bob; and $A\colo R$ \AND\ $B\colo G$.  This implies
a marginal probability of 1/2 for each possibility, $B\colo R$ or $B\colo G$,
for the color of the slip in Bob's envelope. However, upon opening her
envelope and observing that the slip is red, Alice can replace these with the
conditional probabilities $\Pr(B\colo G\vb A\colo R) =1$ and $\Pr(B\colo R\vb
A\colo R) =0$.

\xb\outl{Spin half particles in singlet state sent to Alice and Bob
}\xa

Now suppose that Charlie at the center of a laboratory pushes a button so that
one member of a pair of spin-half particles initially in the singlet spin
state \eqref{eqn34} is sent towards Alice's apparatus at one end of the
building, while the other is simultaneously sent towards Bob's apparatus at
the other end.  If Alice measures the $x$ component of spin of particle $a$
and the outcome corresponds to $S_{ax}=1/2$, what can she say about $S_{bx}$
for particle $b$ traveling towards Bob, assuming that both particles have been
traveling in field-free regions?  By applying the Born rule,
Sec.~\ref{sbct3.3}, using $\ket{\psi_0}$ as a \emph{pre-probability} to a
framework that includes both $S_{ax}$ and $S_{bx}$, Alice, whom we assume is
both a competent experimentalist and has had an up-to-date course in quantum
mechanics, can conclude that $S_{bx}=-1/2$.  And this conclusion is reached by
\emph{precisely the same} sort of statistical reasoning that applies in the
case of colored slips of paper.  Nonlocal influences are involved to no
greater extent than in the case of the colored slips of paper discussed
earlier.

\xb\outl{What happens if Alice changes measurement basis at last moment
}\xa

Suppose on the other hand that Alice decides at the very last minute, after
the two particles are already on their way, to measure $S_{az}$ in place of
$S_{ax}$. How will this change things?  In particular does it somehow alter
the spin of particle $b$, on its way to Bob?  Not at all, as demonstrated by
the detailed analysis in Sec.~23.4 of CQT.  What happens is that Alice learns
something different about her particle, namely the value of $S_{az}$ just
before the measurement takes place.  By employing a different framework,
incompatible with the previous framework, with $S_{az}$ in place of $S_{ax}$,
Alice can now on the basis of the measurement outcome make an inference with
probability 1 about $S_{bz}$, but she loses the ability to say anything about
$S_{bx}$.

\xb\outl{If Alice HAD measured using the other basis? Counterfactual
}\xa

The reader may object that \emph{if} Alice \emph{had} measured $S_{ax}$ rather
than $S_{az}$, she would have gotten a definite value, and from this she
\emph{could have} inferred the value of $S_{ax}$ before the measurement. And
therefore there must have been both a definite value for $S_{az}$ and a
definite value for $S_{ax}$ just before the measurement took place. The words
in italics indicate that the preceding is a counterfactual argument of the
sort philosophers have trouble analyzing. Thus one must approach its use in a
quantum context with particular care; see the discussion in Ch.~19 of CQT.
For present purposes it suffices to note that the conclusion makes assertions
about both $S_x$ and $S_z$ for the same particle at the same time, and thus
violates the single framework rule: we have reached statements which cannot
both be simultaneously embedded in Hilbert space quantum mechanics.  Note that
the issue has to do with \emph{local} properties determined by local
measurements.  Particle $b$ has never been mentioned; it need not even exist.

\xb\outl{Derivations of Bell's inequality: classical locality
}\xa

A more detailed study of derivations of Bell's inequality---we refer the
reader to Ch.~24 of CQT as well as \cite{Grff11,Grff11b}---shows that it is
the matter just referred to, the attempt to ascribe incompatible properties to
a single quantum system, that invalidates derivations of Bell's (or, to be
more precise, the CHSH) inequality when it is applied to the microscopic
quantum world, rather than the macroscopic world of ordinary experience, where
the assumptions needed to justify it can be satisfied, at least to a very good
approximation.  To be sure there is by now an enormous literature devoted to
Bell's inequality, and someone trying to refute its applicability to the
quantum world is in somewhat the same position as the professor challenged by
a student to point out the error in reasoning in an examination paper that has
by a circuitous route arrived at a result which is clearly wrong.  Rather than
seek to identify the error in each and every (supposed) derivation of Bell's
inequality, it seems better to throw the challenge back to the other side.
The nonlocal influence claim violates the principle of Einstein locality
summarized briefly below.  The complete and relatively simple proof is given
in Sec.~6 of \cite{Grff11}. Can the reader find a flaw in it?

\xb
\subsection{Einstein locality}
\label{sbct6.3}
\xa

\xb\outl{Statement of Einstein locality
}\xa

By applying quantum mechanical principles including the single framework rule
one can establish the following principle of \emph{Einstein Locality}.%
\footnote{The wording is essentially the same as in Sec.~2 of \cite{Mrmn98},
  where Mermin refers to it as ``generalized Einstein locality.''  For
  Einstein's own statement see p.~85 of \cite{Enst51}: ``But on one
  supposition we should, in my opinion, absolutely hold fast: the real factual
  situation of the system $S_2$ is independent of what is done with the system
  $S_1$, which is spatially separated from the former.''} %

\begin{quote}

Objective properties of isolated individual systems do not change when 
something is done to another non-interacting system.
\end{quote}

\xb\outl{Hilbert spaces and time developments for systems
}\xa

The proof is given in \cite{Grff11}; what is useful here is a
precise definition of the terms.  Let $A$, Hilbert space $\HM_A$, be the
isolated individual system whose properties are under discussion.  Let $B$
and $C$, Hilbert spaces $\HM_B$ and $\HM_C$, be systems which do not interact
with $A$ during the time interval $t_0$ to $t_1$ of interest in the sense that
the time-development operator factors:
\begin{equation}
  T_{ABC}(t,t') = T_A(t,t')\ot T_{BC}(t,t') 
\label{eqn35} 
\end{equation} 
The role of the ancillary system $C$ is to ``do something'' to $B$; this can
be modeled by varying the initial state $\ket{\phi_0}_C$ of $C$ at $t_0$ while
keeping the initial state of $\ket{\Phi_0}_{AB}$ of system $AB$ remains fixed.
Note that $C$ can interact with, and thus ``do something'' to, $B$ during the
interval from $t_0$ to $t_1$.  The claim that the properties of $A$ do not
change when something is done to $B$ is then established by an argument that
shows that the probabilities associated with a consistent family of
histories of $A$, involving projectors on $\HM_A$ and thus referring to its
properties alone, are independent of the choice made for $\ket{\phi_0}_C$.  In
addition, the consistency of that family does not depend upon
$\ket{\phi_0}_C$.

\xb\outl{Objective properties. Something is DONE another system
}\xa

While the mathematical argument is straightforward, the issue of whether the
English words in the above statement of Einstein locality have been correctly
translated into the mathematics of quantum theory is less so.  Essential to
the argument is, of course, the assumption that any objective properties of
$A$ must be represented by projectors on its Hilbert space, and if the family
of histories contains a sequence of such properties at three or more times
(including the initial time) then the consistency conditions are satisfied.
The properties of $A$ are \emph{objective} in the sense that the whole
situation is modeled from the perspective of a physicist who is \emph{outside}
the system being described.  As always, the sequence of properties under
discussion is determined by the physicist, but any other physicist who uses
the same framework, or a more general one that contains this framework, will
come to exactly the same conclusion about the probabilities.
The next subtlety concerns how the phrase \emph{something is done}, with its
implicit but not insignificant association with the concept of free choice by
a conscious agent, is modeled in quantum terms.  The approach used here, where
a third system $C$ is employed to \emph{do} something to $B$, is at least
consistent with discussions in contemporary quantum information theory where
Alice is said to ``do'' something to a quantum system.  A final issue has to
do with using \emph{initial conditions}. Again, this is consistent with the
approach used in quantum information theory, but buried here is an important
and unresolved problem related to thermodynamic irreversibility; see
Sec.~\ref{sbct8.4}.

Assuming the arguments used to justify Einstein locality are correct, the
result is an extremely simple explanation of why the mysterious nonlocal
influences can carry no information: they do not exist. They are the residue
of a faulty analysis that is inconsistent with the principles of quantum
reasoning needed to resolve the measurement problem(s).

\xb
\section{Quantum Information}
\label{sct7}
\xa

\xb
\subsection{Histories approach}
\label{sbct7.1}
\xa


\xb\outl{Histories answers Bell: Whose info?  Info about what?
}\xa

The histories approach provides a quite definite answer to Bell's
\cite{Bll90} query directed at attempts to explain quantum mysteries by
appeal to information: ``\emph{Whose} information? Information about
\emph{what}?''  The second question belongs to the realm of ontology, and the
ontology which is the subject of the present paper gives a definite answer:
the information is about \emph{quantum properties}, represented mathematically
as subspaces of, or the corresponding projectors on, the quantum Hilbert
space.  More generally, information can be about a time sequence of such
properties, a quantum history.

\xb\outl{Cl information theory uses probabilities
}\xa

\xb\outl{Can be translated to Qm domain as long as single framework is used
}\xa

The modern theory of information, see \cite{CvTh06} is formulated in terms of
probabilities and probabilistic reasoning, the sort found in probability
theory textbooks, and one hopes will someday be used in quantum textbooks.
Thus it is natural to expect that ``classical'' information-theoretic concepts
will appear in quantum information theory in much the same form \emph{as long
  as attention is confined to a single framework}.  This is indeed the case;
e.g., \cite{Grff02}, and while it by no means exhausts the contents of
quantum information theory, it does provide a good beginning, one that is more
satisfactory than the approaches discussed by Timpson in \cite{Tmps08},
which as he himself points out in his Sec.~4.4.1 cannot provide a satisfactory
answer to the question of, as he puts it, ``how things are with a system prior
to measurement.''

\xb\outl{Incompatible frameworks the way to view remaining problems in Qm info
}\xa

Not only does the histories approach provide a good beginning point by
supplying a criterion for how classical information theory can be applied
without fear of generating contradictions or otherwise falling into the
quantum foundations swamp, it also provides a certain perspective on what
remains to be done: the part of quantum information that goes beyond the
classical theory has to do with comparing results that are obtained if one
uses alternative \emph{incompatible} frameworks.  But is not such comparison
forbidden by the single framework rule?  Not at all. What is forbidden is
\emph{combining} the probabilities worked out in different incompatible
frameworks.  Indeed, it is comparison without combining that is resulting in
new physical insights in this very active field of current physics research.
To be sure, the typical practitioner has learned whatever quantum mechanics he
knows from the standard textbooks, and has to make frequent reference to
``measurements'' without any clear idea of what this means.  Research papers
are often filled with complicated mathematical formulas and concepts whose
physical significance is unknown, to the author as well as to the reader.  Bell
would not be pleased.  One can at least hope that a consistent formulation of
quantum ontology will at some future time help bring some order to this
conceptual chaos.

\xb\outl{Whose information? Answered previously in Alice info about particle spin
}\xa

As to Bell's first question, ``whose information?'', the answer has already
been suggested by the little scene presented in Sec.~\ref{sbct6.2}.  The
outcome of her experiment provides Alice with information about the prior
physical state of a particle which her apparatus was designed to measure.  It
is her information, not Bob's information; his particle is completely
unaffected by the distant measurement.  By combining the results of her
experiment with additional information related to the preparation of the
particle pair---see the remarks on preparation in Sec.~\ref{sbct5.2}---she is
then able to infer something about a \emph{microscopic} quantum particle that
is, or was, on its way to Bob, and by this means say something about the
outcome of a measurement which he has already made, or perhaps will make at
some future time.  If the latter, then it is correct to say that Alice has
information about the outcome of a future experiment.  On the other hand, an
ontology which says that the \emph{only} information quantum theory
provides relates to the outcome of future experiments is
seriously inadequate, both for the needs of experimental physicists and for
theorists interested in quantum information; it leaves the black box tightly
closed.

\xb\outl{Additional references to applications in Qm Info
}\xa

Readers with the appropriate technical background who are interested in how an
approach of the sort just discussed can provide a perspective on some problems
in quantum information problems are invited to take a look at \cite{Grff02,
Grff05,LYGG08,CYGG11}.

\xb
\subsection{Sources and information}
\label{sbct7.2}
\xa

\xb\outl{Timpson's claim in Ashgate Companion
}\xa

In Sec.~4.2 of his \cite{Tmps08} Timpson asserts that he can make a
``perfectly precise and adequate definition of quantum information'' based upon
coding ideas.  The strategy is to imagine a source that produces a sequence of
quantum states---it will suffice to consider the case where these are pure
states---analogous to a classical source producing a sequence of symbols drawn
from some alphabet.  A sequence of quantum states is in effect a history of
the sort discussed above in Sec.~\ref{sct2} and could be described in the
language of histories (to which Timpson makes no
reference). However, there is a serious difficulty with this proposal: a
particular quantum state in the sequence is drawn from an alphabet of states
which are \emph{not} required to be orthogonal: see the beginning of his
Sec.~4.2.1.3 (p.~225).  But in this case the history projectors making up the
corresponding family will not in general be orthogonal to each other, so
they will not form a quantum sample space, and considering them to be the
referents of quantum information, what quantum information is \emph{about},
leads into the great swamp.

\xb\outl{Info about records of preparer vs in states themselves
}\xa

\xb\outl{Instrumentalism that Timpson thinks unsatisfactory
}\xa

This has long been appreciated by the orthodox, who, while they would not
state it precisely this way, in effect use the following strategy.  Every time
Alice prepares a state she records the settings on her preparation device.
Distinct macroscopic records correspond to orthogonal quantum states or
projectors, and by tying the records to the not-necessarily-orthogonal
microscopic states of particles prepared by the apparatus one arrives at what
in Ch.~14 of CQT are referred to as dependent or contextual events.  Holevo's
famous bound (see, e.g., Sec.~12.1.1 of \cite{NlCh00}) then tells us the
maximum amount of information about Alice's notebook that Bob's notebook can
contain, following whatever measurement procedure he may employ.  The orthodox
regard this as perfectly sensible, since it makes no reference to what may be
happening inside the black box. If Timpson wants to say that quantum
information is simply information about a macroscopic preparation procedure
his proposal cannot be faulted.  However, by his own standards this amounts to
the sort of instrumentalism he does not find satisfactory as a solution to the
quantum information problem.

\xb
\section{Conclusions }
\label{sct8}
\xa

\xb
\subsection{Summary}
\label{sbct8.1}
\xa

\xb\outl{Qm ontology is realistic. Structure reflected in mathematical theory
}\xa

\xb\outl{Hilbert space description demands a new logic; Qm dynamics indeterministic
}\xa

The consistent ontology for quantum mechanics described in Secs.~\ref{sct2}
and \ref{sct3}, with applications to the classical limit, preparation and
measurement, locality, and quantum information in Secs.~\ref{sct4} to
\ref{sct7}, resembles the ontology of classical mechanics, as represented
mathematically by a phase space and deterministic Hamiltonian dynamics, in
several important respects.  It is realistic: the real world is ``out there'',
not just a part of some observer's consciousness, and its structure is
reflected in the mathematical theory constructed by physicists for describing
it.  But of course it differs from its classical predecessor in important
ways, which can be conveniently summarized under two headings.
First, the Hilbert space description of a system, in which its properties are
represented by subspaces,  requires a new logic in the sense of a
mode of reasoning about the world, with rules somewhat different from those
familiar in classical physics, where ordinary propositional logic fits very
comfortably onto an algebra of physical properties corresponding to subsets of
the phase phase.  Second, quantum dynamics is intrinsically stochastic or
probabilistic: probabilities are present in the basic axioms that apply without
exception to all quantum processes, not just to those associated with some
form of ``measurement.''

\xb\outl{Result: 
}\xa

\xb\outl{No measurement problems
}\xa

\xb\outl{Human consciousness not needed
}\xa

\xb\outl{Objective reality about which two observers can agree
}\xa

\xb\outl{Wave function collapse not needed; use conditional probabilities instead
}\xa

\xb\outl{Nonlocal influences have been banished; no conflict with special relativity
}\xa

\xb\outl{Ontology for quantum info. Shannon works for 1 framework. 
}\xa

\xb\outl{Perspective on remaining Qm info problems
}\xa

By using the new logic and the new dynamics one arrives at an interpretation
of quantum mechanics in which the measurement problems, long the bane of
quantum foundations research, disappear.  Not only does one know how to make
sense of the pointer position at the end of the measurement, one also knows
how to relate it to a property the microscopic system possessed \emph{before}
the measurement took place.  Human consciousness need not be invoked in order
to address or solve these problems, thus getting rid of a difficulty which has
plagued quantum foundations ever since the days of von Neumann's
``psycho-physical parallelism'' \cite{vNmn32b}. Psychology can be cleanly
split off from physics (also see Sec.~\ref{sbct8.4}) to the benefit of both
disciplines.  One can speak of an objective quantum reality: different
observers can agree because there is something ``out there'' in the world,
external to themselves, about which agreement is possible; it is not just a
matter of making bets about the outcomes of future experiments.
Wave function collapse is unnecessary; it can be (and probably should be)
replaced with conditional probabilities, Sec.~\ref{sbct5.2}.  Instantaneous
nonlocal influences have been cleaned out of the quantum foundations swamp,
Sec.~\ref{sct6}, removing an apparent conflict between quantum theory and
special relativity.  It is now possible, Sec.~\ref{sct7}, to provide a
reasonable ontology for quantum information. Its probabilities refer to
quantum properties, and it includes in a natural way, in each single
framework, the standard ideas and intuition of ordinary (Shannon) information
theory. The remaining specifically ``quantum'' problems of information theory
involve comparison of, not combinations of, incompatible frameworks.

\xb\outl{Quantum paradoxes have been resolved
}\xa

In addition to all of this, a whole series of quantum paradoxes
(Bell-Kochen-Specker, Einstein-Podolsky-Rosen, Hardy, \dots) are resolved by
the histories approach.  They have not been discussed in this paper because a
large number have been treated at considerable length in Chs.~20 to 25 of CQT,
and more could be treated by the same methods.  Such paradoxes can be
understood using a consistent set of underlying quantum principles, and not
left as unresolved conundrums, as in \cite{AhRh05}.  Instead they are
interesting, one might even say beautiful, illustrations of ways in which the
quantum world differs from the classical world of our everyday experience, in
much the same way as the twin paradox provides a striking illustration of the
principles of special relativity.

\xb\outl{The following subsections discuss: new logic; Qm dynamics; open problems
}\xa

The new logic and the new (relative to classical mechanics) dynamics, which
form the heart of this paper, are summarized below in Secs.~\ref{sct2} and
\ref{sct3}. The final Sec.~\ref{sbct8.4} indicates some
open problems.

\xb
\subsection{The new logic}
\label{sbct8.2}
\xa

\xb\outl{New logic as radical a break as Copernican solar system
}\xa

\xb\outl{Idea of not combining A, B as A AND B foreign to classical physics
}\xa

The new logic, in which conjunctions and other combinations of incompatible
quantum propositions, represented by noncommuting projectors, are ruled out of
acceptable quantum descriptions by the single framework rule, represents a
radical break with classical physics.  Indeed, one might say it is the central
feature that distinguishes the quantum world from the pre-quantum world of
everyday experience.  It is as radical as Copernicus' shifting the earth out
of the center of the universe, or of Einstein's relativizing time.  The
assumption that if it makes sense to talk about A with reference to some
system, and to talk about B for the same system, then it also makes sense
to talk about A \AND\ B, is essentially automatic in everyday reasoning as
well as in classical physics. A
proposal that this ``obviously true'' fact should not be correct in general for
quantum physics tends to arouse the immediate reaction of ``that cannot make
any sense.''  So it is not surprising that the new logic with its single
framework rule has encountered considerable resistance from those who think it
much too radical to be an acceptable part of good science, even though they
themselves are not able to provide a solution to quantum mysteries by 
other means.  Thus it is important to understand in a clear way where the
single framework rule comes from, what it affirms, and what it does \emph{not}
say.

\xb\outl{Motivation for single framework rule: Qm Hilbert space  + von Neumann
}\xa

\xb\outl{Options for dealing with Hilbert space problems: three approaches
}\xa

\xb\outl{1. Qm logic
}\xa

\xb\outl{2. textbooks invoke measurements
}\xa

First it should be stressed that the single framework rule arises from the
effort to make physical sense of the Hilbert space structure that underlies
all of modern quantum mechanics, as interpreted by von Neumann.  Once the
association of properties with subspaces and their negations with orthogonal
complements of these subspaces has been made, there are serious logical
issues, as Birkhoff and von Neumann pointed out.  How are they to be dealt
with?
Broadly speaking, there have been three approaches.  First, that of quantum
logic, which, as noted in Sec.~\ref{sbct2.1}, has not resolved the conceptual
difficulties of quantum theory, though smarter physicists or even smarter
robots may someday make more progress.  Second, that of standard quantum
mechanics as embodied in the textbooks, where the strategy is not to discuss
the logical issues but instead appeal to measurements.  This reaches its
extreme in the unopenable black box which quantum orthodoxy inserts between
preparation and measurement.  The field of quantum foundations can be thought
of as a protest against this approach of invoking measurement as a way out of
quantum conceptual difficulties.

\xb\outl{3. Histories strategy. Resembles Qm logic, textbooks, orthodoxy in
  certain ways }\xa

\xb\outl{Resembles Qm logic, textbooks, orthodoxy in certain ways
}\xa

The third approach is the histories strategy, in which quantum descriptions of
the world are split off into families or frameworks that cannot be combined
with each other, but within each framework ordinary propositional logic
applies, along with the usual rules and intuition associated therewith.  The
histories formulation has some things in common with the other two approaches.
It is a form of quantum logic in the sense of a scheme for correct reasoning in
the quantum domain that differs from the logical scheme of classical physics.
However, its single framework rule is not part of what is usually referred to
as quantum logic.  And it shares with textbook quantum mechanics and quantum
orthodoxy the refusal to talk about certain things.  But what it does allow the
physicist to talk about is now greatly enlarged: it includes a whole series of
microscopic properties and events. Including a significant set of things that
experimental physicists think they are able to detect with their instruments.
For these reasons the logic used in histories is perhaps not quite as radical
or as innovative as might at first be supposed, though it obviously includes
some very new features.

\xb\outl{Serious misunderstandings of histories approach
}\xa

\xb\outl{1. Single framework rule not to be taken seriously
}\xa

There have, nonetheless been some severe misunderstandings of 
the histories approach to this problem by means of its single framework rule.
A first type of misunderstanding arises from supposing that the single
framework rule, perhaps because it is both unfamiliar and not presented in the
textbooks, does not have to be taken seriously.  Claims that the histories
approach leads to contradictions, as in \cite{Knt98,BsGh99,BsGh00}, can be
refuted, as in \cite{Grff98,GrHr98,Grff00,Grff00b}, by working through the
argument and seeing at which point the claimant has, perhaps unwittingly,
strayed from one framework onto a different, incompatible one in the course of
constructing a logical argument.  While what is going on is clearest in
situations like the Bell-Kochen-Specker paradox, there are also situations in
which the consistency conditions of quantum dynamics play an important role,
as in the case of the Aharonov and Vaidman three box paradox \cite{AhVd91}
which forms the basis of Kent's criticism in \cite{Knt98}; see the detailed
discussion of this paradox and the associated incompatible frameworks in
Sec.~22.5 of CQT.

\xb\outl{Histories approach does not lead to contradictions
}\xa

It is worth noting at this point that there is a a very general argument,
Ch.~16 of CQT, that the histories approach will not result in contradictions,
and critics have yet to find any flaws in it.  To be sure there may be flaws
that lie undiscovered because the logical structure employed in the histories
approach has yet to be subjected to sufficiently severe scrutiny by those who
have first taken the trouble to carefully understand what it is all about. The
author hopes the present paper, by addressing various misunderstandings in a
direct way, may provoke that sort of serious study of a system of quantum
interpretation which, on the basis of the problems and paradoxes it resolves,
has some claim to being the best and most consistent approach to quantum
ontology currently available.

\xb\outl{2. Incompatible frameworks vs. mutually exclusive
}\xa

\xb\outl{Dowker and Kent, Wallace are instances of this misunderstanding
}\xa

A second type of misunderstanding is to suppose that the incompatibility of
frameworks, which prevents their being combined, is the same as their being
\emph{mutually exclusive}, one is true and the other is false, in the same
sense that the events of a coin landing heads or tails are mutually exclusive.
This leads among other things to the erroneous idea that histories quantum
mechanics is incomplete unless it includes some ``law of nature'' that
specifies the correct framework that obtains in a particular physical
situation.  A particular case is the Dowker and Kent \cite{DwKn96} critique of
the notion of a quasiclassical framework discussed in Sec.~\ref{sbct4.2}.
Another instance is provided by Wallace's assertion (p. 39 in \cite{Wllc08})
that the histories approach leads to a view of reality that only makes sense
``when described from one of indefinitely many contradictory perspectives.''
Presumably these ``perspectives'' are frameworks; see the refutation in
\cite{GrHr98} of a similar claim in \cite{Knt98}.

\xb\outl{Incompatible vs mutually exclusive: mixup due to use of measurements
}\xa

One possible source for the mixup between the relationship of quantum
incompatibility and that of being mutually exclusive is the unfortunate
reliance in textbooks upon an unanalyzed (and, for the orthodox, unanalyzable)
measurement process when presenting a physical interpretation of quantum
mechanics.  The books state, correctly, that there is no measurement which can
simultaneously determine $S_x$ and $S_z$ for a spin half particle.  The
measurement setups required in these two cases \emph{are} mutually exclusive:
they correspond to macroscopically distinct arrangements, so the subspaces of
the quantum Hilbert space required to represent them are necessarily
orthogonal to each other.  That does not of course mean that the microscopic
properties measured by these distinct apparatus setups are also mutually
exclusive.

\xb\outl{Ordinary logic in the macroscopic world
}\xa

If the fundamental logic needed to describe the quantum world is so different
from the logic of everyday affairs, why is the latter so successful and
widespread?  The histories approach answers this question in the manner
sketched in Sec.~\ref{sct4}: to describe the macroscopic world one only needs
a single quasiclassical framework or, to be more precise, all frameworks of
this type give the same results ``for all practical purposes.''  Until one
arrives at situations in which quantum effects (those associated with
noncommuting operators on the Hilbert space) become important, classical
mechanics is adequate, and one would expect that whenever classical
mechanics is adequate the logic that corresponds so well to subsets of the
phase space will also be adequate.

\xb
\outl{Analogies help, but serious student needs to consider examples}
\xa

 Obviously the single framework rule is
not something easy to understand.  The author hopes that the various
analogies and comparisons given above in Sec.~\ref{sbct2.3} will assist the
reader in gaining some intuitive grasp on an unfamiliar concept.  However, for
the serious student of quantum mechanics there is no substitute for working
through various specific quantum examples, such as the those scattered
throughout CQT.

\xb
\subsection{Indeterministic dynamics}
\label{sbct8.3}
\xa

\xb\outl{Histories time development is stochastic
}\xa

\xb\outl{Histories Hilbert space, sample space
}\xa

The histories approach assumes that quantum time development is basically
stochastic or probabilistic.  Always, not just when measurements are being
carried out. One starts with a history Hilbert space $\breve\HM$, a tensor
product of the sample space of a single system at each of the (discrete, and,
for convenience, finite in number) times of interest. Elements of the sample
space, referred to as histories, are then projectors forming a PD of the
identity on $\breve\HM$.  As in other applications of probability theory to
stochastic processes, the physical (ontological) interpretation is that just
one of the possible histories in a given sample space actually occurs.  In
this sense the histories approach is, as noted in Sec.~\ref{sbct3.5}, quite
distinct from many-worlds versions of quantum theory in which things that do
not occur here are supposed to take place in some alternative but inaccessible
universe.

\xb\outl{Closed quantum system: weights if consistency conditions satisfied
}\xa

\xb\outl{Consistent family. Compatible families: refinement must be consistent
}\xa

\xb\outl{Single framework rule means incompatible families cannot be combined
}\xa

In the case of a closed quantum system weights can
be assigned to elements of the sample space provided a consistency condition,
vanishing of the  off-diagonal elements of the decoherence functional,
are satisfied, and a consistent set of probabilities defined using the weights
and additional data (such as an initial state).  The Born rule is a particular
example of this scheme when only two times are involved, in which case
the consistency conditions are always satisfied.
Provided the consistency conditions are satisfied such a sample space, and its
associated event algebra, forms a consistent family or framework.  Two such
frameworks can be combined only if the different projectors commute (the same
rule as for systems at a single time) and if the common refinement resulting
from products of these projectors also satisfies the consistency condition.
Otherwise they are incompatible, and the single framework rule forbids
combining them for purposes of calculating probabilities or carrying out
probabilistic reasoning of the sort that reduces to ordinary propositional
logic when probabilities are 0 and 1.

\xb\outl{Probabilistic dynamics more acceptable than new logic
}\xa

Because stochastic models are widely used in many branches of science,
including physics, and because probabilities are already found, though not
very well explained, in quantum textbooks, the second major way in which the
histories quantum mechanics departs from classical mechanics, its use of
probabilistic time development as a fundamental law, not just for measurements
or macroscopic systems, may be a bit easier to accept than the new logic.
Nonetheless, the notion that time development is fundamentally indeterministic
is one which many physicists find unappealing, even if not completely
unacceptable.  The objections by Einstein are well known, and the continuing
popularity, at least in some quarters, of the Everett and Bohm interpretations
indicates something of the appeal of classical determinism.  To be sure, one
can speculate that even Einstein might have found stochastic quantum time
development acceptable if one consequence would be to make it consistent (in a
sense he might not have anticipated) with his notion of locality,
Sec.~\ref{sbct6.3}.

\xb\outl{Consistency conditions further restrict range of meaningful statements
}\xa

Note that the general strategy involved in the new logic, that of restricting
the domain of logical or, in this case, probabilistic reasoning by a rule that
prevents combining incompatible frameworks, is central to the histories
discussion of time dependence.  When assigning probabilities to a closed
quantum system not only must one use a PD of the history identity operator,
but an additional consistency condition must be fulfilled for histories
involving more than three times, Sec.~\ref{sct3}.  Whereas these consistency
conditions can be stated in a very simple and clean mathematical form,
\eqref{eqn15}, they still look a bit strange, and have no (obvious)
counterpart in the notions needed to describe a quantum system at a single
time.  While no inadequacy has been identified in the current histories
formulation of stochastic time development, there may be something more
interesting lurking there; see below.

\xb
\subsection{Open issues}
\label{sbct8.4}
\xa

\xb\outl{Ontology uses: Hilbert space (no hidden variables); prob. time development
}\xa

\xb\outl{No measurement problems. Qm world is local. Gives basis for Qm info.
}\xa

The quantum ontology introduced in this paper seems satisfactory in a number
of respects.  It is based upon the quantum Hilbert space so has no need for
additional (hidden) variables.  Time development is stochastic, and
calculating the probabilities requires only the use of unitary time
development induced by Schr\"odinger's equation, not some modification
thereof.  There is no measurement problem: measurements (and preparations) can
be consistently understood using the same tools employed for all other quantum
processes, and measurement apparatus designed by competent experimentalists
does what it was designed to do: reveal properties microscopic systems had
\emph{before} the measurement took place.  The quantum world is \emph{local}
when allowance is made for the width of quantum wave packets. There is neither
action nor passion at a distance. And one has a preliminary, but thus far
satisfactory, ontology for quantum information.

\xb\outl{Many problems remain to be solved. E.g. nanoscale electronic transport
}\xa

That is not to say all problems in the domain of quantum interpretation have
been solved.  Some of the work that still needs to be done is quite technical;
e.g., developing appropriate conceptual tools to describe electronic transport
in nanostructures, in order to replace the present quasiclassical descriptions
adequate for larger structures but inadequate at the smallest scales.
Whatever the practical importance of such work, it is unlikely to appeal as a
subject of research to philosophers and physicists concerned with foundations
of quantum theory.  But there are other topics they should find more
interesting. Here are four of them.

\xb
\subsubsection{Entangled histories}
\label{sbct8.4.1}
\xa

\xb\outl{Entangled histories
}\xa

Whereas the principles of quantum stochastic dynamics summarized in
Sec.~\ref{sct3} are both consistent and provide what seems to be a quite
adequate foundation for all the sorts of calculations taught in textbooks and
used in current research papers, they are incomplete in the following sense.
Most discussions of histories and all discussions of consistency conditions
known to the author employ a sample space of \emph{product} histories: at each
time in the history tensor product space a projector represents a property of
the system \emph{at that particular time}.  But the tensor product space
representing a composite quantum system---two or more subsystems---at a single
time also contains what are called \emph{entangled states}, which cannot be
thought of as assigning a particular property to each subsystem; e.g., the
singlet state \eqref{eqn34}.  Consequently, the tensor product space of
histories also includes states which are, so-to-speak, entangled between two
or more times.  What is their physical significance?  Could they serve a
useful role in describing some sort of interesting time development?  And how,
assuming it to be possible, are probabilities to be assigned in the case of a
closed quantum system?  It is not obvious how consistency conditions as
presently formulated, see \eqref{eqn15}, can be extended to this case, since
the temporal ordering of events plays a crucial role.  Thus this is thus an
open question.

\xb
\subsubsection{Sufficiency of the language}
\label{sbct8.4.2}
\xa

\xb\outl{Single framework rule restrictive: does it allow enough?
}\xa

\xb\outl{At present the answer is ``yes''
}\xa

There is an obvious question about an approach to quantum interpretation
which employs the histories strategy as embodied in the single framework rule,
and thus declares various subjects out of bounds because they involve
meaningless combinations of incompatible frameworks. Are there important topics,
topics central to understanding the quantum mechanical world, which are
thereby excluded?  Or, to put the matter a different way: is there sufficient
flexibility in the language of quantum mechanics, when subject to the single
framework rule, to allow the sorts of descriptions and modes of reasoning
which in practice quantum physicists need in order to pursue their discipline?
At the present time the answer seems to be ``yes.''  But this is
necessarily tentative, for quantum mechanics is by now a vast discipline
with an enormous range of applications, and no individual can be expected to
be familiar with all of them. The fact that no failure has come to light
in the decade following the publication of CQT lends support to the
idea that the histories approach is sufficient.  Its success in terms of
treating numerous toy models and resolving numerous paradoxes suggests
that exceptions to its rules, if they exist, may be hard to find.

\xb\outl{Examples suggest 'single framework' draws the line at the right place
}\xa

\xb\outl{Question of adequacy remains open.  Critical examination is welcome
}\xa

Indeed, the plethora of examples studied thus far would suggest that the
single framework rule ``draws the line'' rather much in accord with the 
quantum physicist's intuition.  The mysterious nonlocal influences that cannot
carry information, and are thus forever beyond the reach of experimental
evidence, disappear when the single framework rule is enforced,
Sec.~\ref{sct6}.  On the other hand, the black box of quantum orthodoxy has
been pried open, and the belief of experimental physicists that the apparatus
they build really measures something has been confirmed: they have been right
all along to ignore the strictures found in the books from which they first
learned quantum theory.  But there are also experiments involving
interference effects in which one must be more careful---and the histories
approach not only indicates which these are, but suggests useful ways to
think about them.
Nevertheless, the question of adequacy remains open, and one can certainly
imagine the possibility that further research will uncover defects in the
histories approach which have hitherto escaped the attention of its advocates.
A critical examination of histories ideas and conclusions by those who have
taken the trouble to try and understand them would be most welcome; one aim of
the present article is to encourage such.

\xb
\subsubsection{Thermodynamic irreversibility}
\label{sbct8.4.3}
\xa

\xb\outl{Standard QM will not work: measurements are irreversible
}\xa

A problem belonging to a somewhat different category is that of understanding
thermodynamic irreversibility.  Frigg's discussion in \cite{Frgg08} is
limited to \emph{classical} statistical mechanics, and this for a very good
reason. Little progress can be expected in studying thermodynamic
irreversibility from a fundamental quantum perspective as long as one has to
depend upon the textbook formulation using measurements.  Everyone agrees that
measurements are thermodynamically irreversible processes.  Thus a discussion
that relies on them must introduce at the outset, in an arbitrary and totally
uncontrolled manner, the very phenomenon one is trying to understand. That is
not very appealing.

\xb\outl{Histories approach has no measurement problem
}\xa

\xb\outl{Time reversal invariance (or lack thereof) is not an issue
}\xa

By contrast, the histories formulation has no measurement problem.  But it
does introduce probabilities at a fundamental level; is this equally bad?  No,
for even classical probabilistic descriptions \emph{need not} result in
assigning a direction to time; see \cite{Frgg08,Uffn10}.  That the
histories probability assignment in fact does \emph{not} introduce a time
direction is evident in its fundamental formulas used to check for consistency
and assign probabilities: \eqref{eqn15} and \eqref{eqn16} in
Sec.~\ref{sbct3.4}. (Note that taking the adjoint of the chain operator
$K(Y^\al)$ amounts to reversing the order of the times.)
It is necessary at this point to clear up a matter that perennially causes
confusion.  Unitary time development means that $T(t',t) = T(t,t')\ad$ is the
inverse quantum operator to $T(t,t')$, and this is what is behind the assertion
that the histories approach does not single out a direction of time. However,
that is \emph{not} the same as saying that the (closed) quantum system under
discussion is invariant under the symmetry operation of \emph{time reversal},
which amounts to imposing an \emph{additional} condition connecting $T(t',t)$
and $T(t,t')$.  (That time reversal invariance in this sense is not relevant to
understanding the second law of thermodynamics follows from the observation
that systems placed in magnetic fields remain irreversible.)

\xb\outl{Way now open for discussion of irreversibility in consistent Qm terms
}\xa

With the measurement problem disposed of, the way is now open to develop an
understanding of thermodynamic irreversibility in consistent \emph{quantum}
terms. This may or may not clear up any of the conceptual difficulties which
arise in the classical case; that remains to be seen. But it has become a
a reasonable and interesting topic for research.

\xb
\subsubsection{Epistemology}
\label{sbct8.4.4}
\xa

\xb\outl{Objective approach. QM not needed to discuss consciousness 
}\xa

A fourth problem concerns epistemology: how do human beings know things and
what is the relationship of this (supposed) knowledge to the real world,
especially the microscopic quantum world?  The histories ontology does not
automatically provide an answer to this question.  It does provide a framework
(now using that word in a nontechnical sense) for rational discussion and
exploration, one that is different from classical mechanics, but not so
different as to make useless the insights provided by classical physics.
Sensible quantum descriptions can be constructed from the perspective of
someone outside the system being considered.  It makes quantum sense to speak
of neurons in the brain in the manner of physiologists, and to imagine
memories and consciousness itself as somehow represented or expressed in an
appropriate quantum framework, whether or not things are being measured or
observed in some way.  Consciousness, brain function, etc.\ remain problems to
be solved, but quantum mechanics need not be essential to their formulation or
discussion, provided there is a suitable quasiclassical framework within which
the quantum world can exhibit itself in the familiar classical terms of
signals, perhaps with stochastic corrections, traveling along neurons, etc.
If nothing else, clearing away the cobwebs of wave function collapse, many
minds, nonlocal influences, and the like may free up for the serious study of
important epistemological problems some of the intellectual power that might
otherwise be dissipated in the quantum swamp.

\xb
\section{Acknowledgments}
\xa

The research
described here received support from the National Science Foundation through
Grants 0757251 and 1068331.


\xb
\end{document}